\documentclass[twocolumn,10pt]{article}

\usepackage[utf8]{inputenc}
\usepackage[T1]{fontenc}
\usepackage{lmodern}
\usepackage[margin=0.85in]{geometry}
\usepackage{graphicx}
\usepackage{booktabs}
\usepackage{amsmath}
\usepackage[numbers,sort&compress]{natbib}
\usepackage{xcolor}
\usepackage[colorlinks=true,allcolors=blue]{hyperref}
\usepackage{dblfloatfix}

\usepackage{listings}
\graphicspath{{experiments/results/}{figures/}}

\newcommand{\atlas}[1]{%
  \IfFileExists{experiments/results/#1.png}%
    {\includegraphics[width=0.31\textwidth]{experiments/results/#1.png}}%
    {\fbox{\parbox[c][3cm][c]{0.29\textwidth}{\centering\footnotesize
       \textcolor{red}{\texttt{\detokenize{#1}}}\\[2pt](panel not generated)}}}%
}

\title{\bf Mapping the Narrow Corridor with Large Language Models}

\author{Ebrahim M. Songhori \\ \texttt{e.songhori@gmail.com} \\[3pt]
  {\small Code: \url{https://github.com/esonghori/narrow-corridor-llm}} \\
  {\small Gallery: \url{https://esonghori.com/narrow-corridor-llm/gallery/}}}
\date{July 1, 2026}

\begin{document}
\maketitle

\begin{abstract}
Acemoglu and Robinson (2024 Nobel laureates in Economics) trace national
histories in \emph{The Narrow Corridor} as paths through a two-dimensional
(state power, society power) space but plot none: quantifying the two axes at
each moment is the labor-intensive,
judgment-laden task that has kept the framework qualitative. We ask whether large
language models (LLMs) can operationalize it. We introduce a reproducible,
provider-agnostic pipeline that scores a country period by period using
chain-of-thought (events first, then score) and in-context anchoring on prior
periods, against an explicit rubric with schema-validated output. As a proof of
concept we produce a twelve-country trajectory atlas (Iran, France, the United
Kingdom, the United States, China, Chile, Colombia, the Democratic Republic
of the Congo, Lebanon, Zambia, Somalia, and India), chosen to place at least two
countries in each of the book's four
Leviathan types, and compare four models. We assess the
scores for consistency with an expert index (V-Dem) and for inter-model
agreement, reading agreement as concurrent consistency rather than accuracy,
since such indices likely appear in the models' training data, and discuss
conceptually how the biases of an LLM annotator and a human expert differ. Code,
prompts, runs, and an interactive gallery of the animated trajectories are
released.
\end{abstract}

\section{Introduction}
\label{sec:intro}

In \emph{The Narrow Corridor}~\citep{acemoglu2019narrow}, Acemoglu and Robinson
(awarded the 2024 Nobel Prize in Economics in part for this line of institutional
research) propose that liberty is neither the natural product of a strong state nor of a
strong society, but of a contested balance between the two. They picture a
country's history as a path through a plane whose horizontal axis is the power
of civil society and whose vertical axis is the power of the state; a ``narrow
corridor'' running between the axes marks the region where the two powers grow
together and liberty is sustained. The book is rich in narrative but contains
\emph{no plotted trajectories}. The obstacle is not conceptual but
operational: turning centuries of a nation's history into a sequence of
$(\text{society}, \text{state})$ coordinates requires an enormous amount of
expert judgment, and any such coding is vulnerable to the biases, selective
readings, and coarse simplifications that historians rightly distrust.

This is a measurement problem, and measurement problems in the social sciences
increasingly admit machine assistance. Large language models have been shown to
annotate and classify political text at or above the level of crowd
workers~\citep{gilardi2023chatgpt}, and a growing literature examines their use
as instruments in computational social science~\citep{ziems2024can}. We ask a
narrower, concrete question: \emph{can an LLM place a country in the Narrow
Corridor space, period by period, in a way that is reproducible, checkable
against an established expert index, and useful to a researcher?}

We take an applied, digital-humanities stance. Our primary artifact is an
\textbf{twelve-country trajectory atlas} (Iran, France, the United Kingdom, the
United States, China, Chile, Colombia, the Democratic Republic of the Congo,
Lebanon, Zambia, Somalia, and India), rendered as time-shaded paths through the
corridor, together with the cross-country patterns those paths reveal. The twelve
are chosen so that at least two countries fall in each of the book's four
Leviathan types (Section~\ref{sec:setup}), which lets us ask whether an LLM
places each country in the region the book assigns it. The
method and its validation are the backing that makes the atlas trustworthy
rather than decorative. Concretely, we contribute:

\begin{enumerate}
  \item \textbf{A scoring method.} A period-by-period pipeline combining an
        anchored $0$--$10$ rubric, chain-of-thought elicitation of events and
        trends~\citep{wei2022chain}, and in-context anchoring on prior
        periods~\citep{brown2020language,dong2024survey}.
  \item \textbf{A provider-agnostic, open implementation.} Any model reachable
        through a unified interface can be swapped by changing one string; we
        release code, prompts, caches, and every run.
  \item \textbf{A trajectory atlas and cross-country reading} of twelve countries
        spanning the book's four Leviathan types.
  \item \textbf{An interactive, publicly hosted web gallery.} A static site
        (GitHub Pages) presenting every trajectory as a time-shaded plot with
        \emph{click-to-play animations} that reveal the path period by period
        and annotate the historical event driving each move; readers explore and
        compare models directly in the browser, turning a static figure into a
        navigable atlas.
  \item \textbf{An assessment of the scores} for consistency with an expert
        index (V-Dem) and for inter-model reliability, with explicit treatment of
        the training-data-overlap threat that makes such agreement concurrent
        consistency rather than proof of accuracy.
  \item \textbf{A conceptual discussion of measurement bias}: what an LLM
        annotator plausibly gets right and wrong relative to a human researcher
        or an expert panel doing the identical task. We do not commission a fresh
        panel of coders; V-Dem is itself exactly such a study, a large expert
        coding operation with many raters per case, so our comparison against it
        (Section~\ref{sec:validation}) already stands in for the human-expert
        baseline.
\end{enumerate}

\section{Related Work}
\label{sec:related}

\paragraph{The Narrow Corridor and measuring state/society power.}
The (state, society) framing extends the institutional program of Acemoglu and
Robinson~\citep{acemoglu2012why,acemoglu2019narrow}. Quantifying the two axes
connects to decades of effort to measure regime characteristics and state
capacity: the Varieties of Democracy (V-Dem) project aggregates expert codings
into latent indices including a core civil-society index and measures of state
administrative capacity~\citep{coppedge2024vdem,pemstein2018vdem}; Polity5 codes
regime authority on an autocracy--democracy scale~\citep{marshall2020polity};
and Freedom House rates political rights and civil
liberties~\citep{freedomhouse2023}. These projects are themselves large expert
coding operations, which makes them both our natural validation targets
(Section~\ref{sec:validation}) and our natural comparison point when we ask what
an LLM changes about the coding process (Section~\ref{sec:discussion}).

\paragraph{Text as data and LLMs as social-science instruments.}
The use of automated methods to extract measures from text is
well established~\citep{grimmer2013text}. Recent work reports that instruction-tuned
LLMs match or exceed crowd annotators on political text-labeling
tasks~\citep{gilardi2023chatgpt} and surveys their broader promise and pitfalls
for computational social science~\citep{ziems2024can}. A related strand uses LLMs
to \emph{simulate} human populations~\citep{argyle2023out} and debates whether
they can stand in for human participants at
all~\citep{dillion2023can}. Our task differs from labeling or simulation: we ask
the model to produce a continuous, temporally coherent \emph{measurement} of a
theoretical construct, then check that measurement against independent data.

\paragraph{Prompting.}
Our pipeline relies on two now-standard techniques. Chain-of-thought
prompting~\citep{wei2022chain} improves reasoning by eliciting intermediate
steps; we use it to force the model to enumerate historical events and trends
\emph{before} scoring. In-context learning~\citep{brown2020language,dong2024survey}
lets the model condition on examples provided at inference time; we feed all
prior periods' scores forward so the trajectory is temporally coherent rather
than a sequence of independent guesses. This anchoring is a deliberate design
choice with a cost (an early mis-score can propagate) that we return to in
Section~\ref{sec:limitations}.

\paragraph{Bias and reliability.}
LLMs inherit biases from training data~\citep{bender2021dangers} and exhibit
measurable political leanings: they do not reflect a representative cross-section
of opinion~\citep{santurkar2023whose}, can carry the political slant of their
pretraining corpora into downstream behavior~\citep{feng2023pretraining}, and
have been probed directly for partisan
bias~\citep{motoki2024more}. These findings motivate both our reliability
analysis, which borrows the logic of inter-coder reliability from content
analysis~\citep{krippendorff2018content}, and our discussion of how LLM bias
compares to the well-known biases of human expert coders.

\section{Method}
\label{sec:method}

\begin{figure}[t]
  \centering
  \includegraphics[width=\columnwidth]{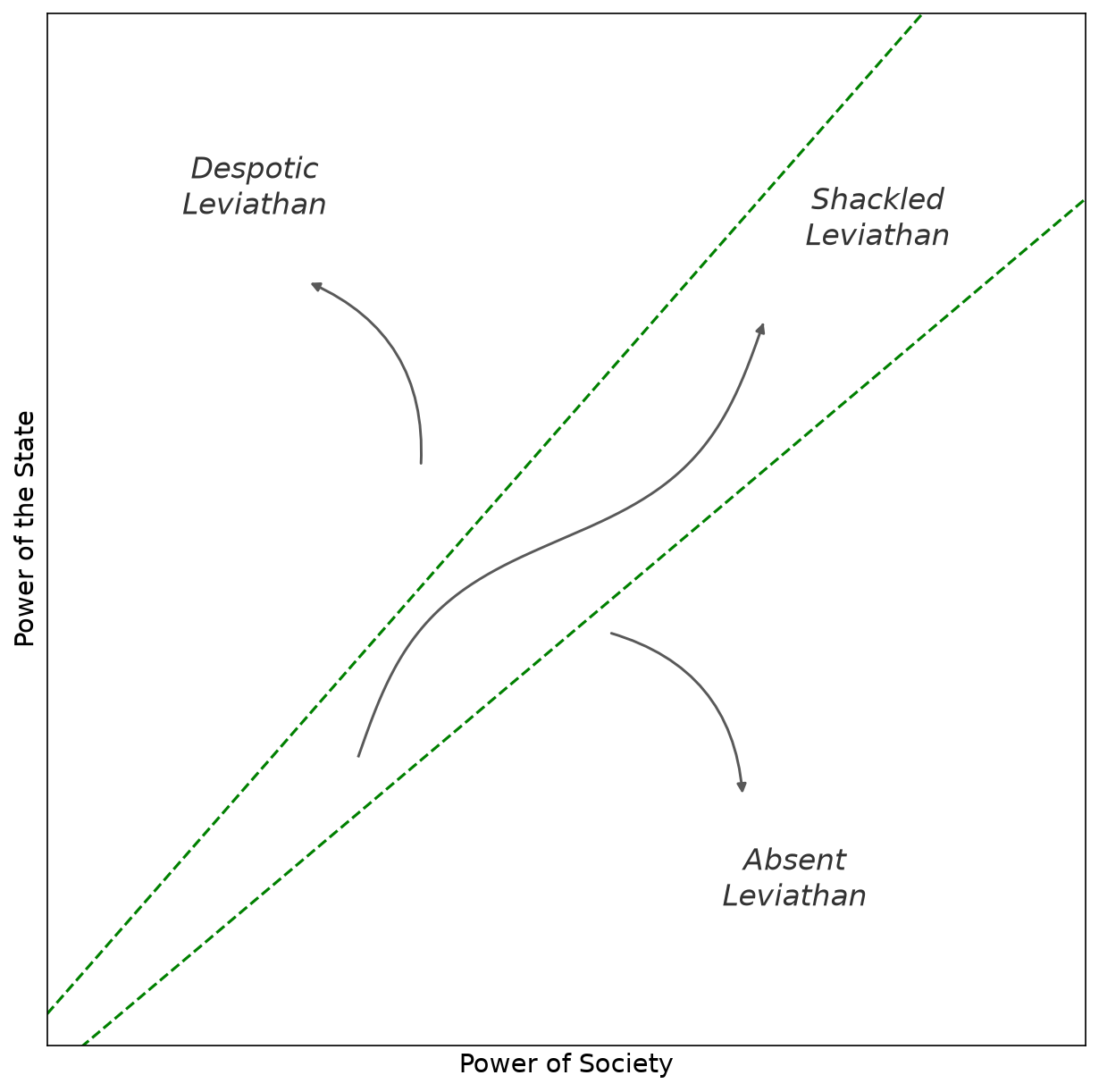}
  \caption{The Narrow Corridor~\citep{acemoglu2019narrow}. The dashed lines are
  the book's \emph{widening} corridor; a country whose path stays inside it is a
  Shackled Leviathan, one above it a Despotic Leviathan, and one in the
  weak-state region below it an Absent Leviathan.}
  \label{fig:corridor}
\end{figure}

\paragraph{The space.} We place a country at each time period $t$ at a point
$\big(\mathrm{soc}_t, \mathrm{sta}_t\big) \in [0,10]^2$, where $\mathrm{soc}_t$
is the power of civil society and $\mathrm{sta}_t$ is the power of the state. A
period is a fixed span of years (e.g., a decade). We read a country as
\emph{in the corridor} in period $t$ when its state and society power are close
(near the diagonal) and \emph{above} or \emph{below} it when one clearly
dominates. Following the book, the corridor in our plots is \emph{widening}: the
dashed boundaries are narrow where state and society are both weak (bottom-left)
and diverge as the two powers grow together (top-right), because liberty becomes
easier to sustain once both are strong. They mark the band illustratively, not as
a fitted threshold, so positions are read qualitatively, from where a path sits
relative to the widening band, not from a numeric cutoff.
Figure~\ref{fig:corridor} sketches the space and the book's four Leviathan
regions in this style.

\paragraph{Anchored rubric.} A free-floating request for ``two numbers'' is not
a measurement. Every scoring prompt embeds a fixed $0$--$10$ rubric that defines
each band of state power (from $0$--$2$, collapsed/nonexistent, to $9$--$10$,
overwhelming penetration of society) and society power (from $0$--$2$, atomized,
to $9$--$10$, society pervasively checks the state), plus explicit neutrality
guidance: judge each period on its own terms, weigh evidence for and against a
shift, avoid \emph{presentism} (scoring a period by how things later turned out,
or by today's standards, rather than by the conditions and information of the
period itself), and treat the two axes as independent. The rubric is
what makes scores comparable across periods, countries, and models.

\paragraph{Chain-of-thought: events then scores.} For each period the model is
first asked to list the major events and trends affecting state or society power,
separating slow-moving trends from discrete events and noting the direction of
each effect. This narrative is then inserted into a second prompt that requests
the numeric score, so the number is conditioned on explicit reasoning rather
than produced in one leap~\citep{wei2022chain}.

\paragraph{In-context learning: temporal anchoring.} Starting from the earliest
period (scored on absolute terms), we accumulate every prior period's
$(\mathrm{soc},\mathrm{sta})$ pair and feed the running trajectory into each
subsequent scoring prompt~\citep{brown2020language}. The model reports both the
change from the previous period and the resulting absolute values, which
discourages the sequence from drifting and keeps a quiet period near zero
change.

\paragraph{Structured, validated output.} Each scoring call returns a JSON
object (the two power values, their per-period changes, a short key event, and
a one-sentence justification) validated against a fixed schema. A malformed
response triggers one reminder retry and then a numeric fallback extractor before
an error is raised.

\paragraph{Provider-agnostic access.} All model calls go through a
single unified interface, so a model is selected by one identifier string and
providers can be swapped without code changes. This is what lets us run the
identical prompts across four model families and isolate model effects from
prompt effects.

\paragraph{Visualization.} A run renders as (i) a static plot of society vs.\
state power with the path time-shaded from early (dark) to late (bright) and a
year colorbar, the largest moves annotated with their year and key event; and
(ii) an animation that reveals the path period by period, drawing an arrow for
the move each event caused and captioning it. Figure~\ref{fig:atlas} is built
from the static renderings; the animations are best viewed in the interactive
gallery at \url{https://esonghori.com/narrow-corridor-llm/gallery/}.

\section{Experimental Setup}
\label{sec:setup}

\paragraph{Countries.} We study twelve countries, chosen so that at least two
fall in each of the four Leviathan types the book defines. \textbf{China} and
\textbf{Iran (Persia)} are candidate
\emph{Despotic} Leviathans (a strong state over a weak society); the
\textbf{United Kingdom}, \textbf{France}, the \textbf{United States}, and
\textbf{India} are candidate \emph{Shackled} Leviathans (state and society
balanced inside the corridor, India's under the ``cage of norms'' the book
stresses); \textbf{Colombia}, \textbf{Zambia (Northern Rhodesia)}, and
\textbf{Lebanon} are candidate \emph{Paper} Leviathans (a state that exists on
paper, a constitution and formal institutions, but cannot enforce its authority);
and the \textbf{Democratic Republic of the Congo} and \textbf{Somalia} are
candidate \emph{Absent} Leviathans (a state that is effectively absent, lacking
even the forms of one). \textbf{Chile} is included as a leave-and-re-enter case
around the 1973 coup. Crucially, the Paper and Absent types are \emph{not}
distinguished by position in the (society, state) plane, both are weak states low
on the state axis, but \emph{de jure}: a Paper Leviathan has the forms of a state
it cannot enforce (Lebanon), an Absent Leviathan lacks even those (Somalia). Our
figures therefore label only the three spatially distinct regions (Despotic
above the corridor, Shackled inside it, and the weak-state region below); the
Paper/Absent distinction is drawn here in the text. Labeling each a
``candidate'' is deliberate: whether the LLM actually places a country in its
assigned region, and in the corridor's expected part, is a question we test, not
an assumption (Section~\ref{sec:atlas}). Each country is scored from a country-appropriate
start year through 2020 (a common cutoff for consistency across cases) in decade
(10-year) periods, giving 11--24 periods per country
(Table~\ref{tab:validation} lists windows and counts). Twelve countries is a
proof of concept, not a basis for cross-national generalization.

\paragraph{Models.} We compare four models from four families (Google,
Anthropic, OpenAI, Alibaba): three closed-weight models (\texttt{gemini-2.5-pro},
\texttt{claude-opus-4-8}, \texttt{gpt-5.5}) and one open-weight model
(\texttt{qwen-2.5-72b-instruct}), which keeps an open- vs.\ closed-weight
contrast. All are driven through the same interface with identical prompts,
isolating model effects from prompt effects.

\paragraph{Consistency check against an expert index.} We compare the LLM scores
to V-Dem~\citep{coppedge2024vdem}, aligning each period's midpoint year to the
V-Dem year. We map LLM \emph{society power} to the V-Dem core civil-society index
(\texttt{v2xcs\_ccsi}) and LLM \emph{state power} to a state-authority measure
(V-Dem state authority over territory, \texttt{v2svstterr}), a narrow,
territorial-control proxy for the rubric's broader penetration-of-society notion
of state power. The mapping is fixed before inspecting correlations, and
alternative capacity indicators are easy to swap in the released code. We report
the \emph{Spearman rank correlation} $\rho$ per country, a number from $-1$ to
$+1$ that measures whether two series move up and down together in the same
rank order ($+1$ perfect agreement, $0$ none, $-1$ opposite), computed on both
\emph{levels} and \emph{first differences} (period-to-period change): levels are
dominated by a shared secular trend and inflate the correlation, so first
differences are the more honest agreement signal. Crucially, V-Dem is
itself almost certainly in the models' training data, so we read this as
\emph{concurrent consistency with expert consensus}, not as external proof of
accuracy (Section~\ref{sec:limitations}). Polity5 and Freedom House are natural
additional anchors but are out of scope here.

\section{Results}
\label{sec:results}

\subsection{Trajectory atlas}
\label{sec:atlas}
\begin{figure*}[p]
  \centering
  % Panels match run_experiments.py slugs: <country-slug>__<model-slug>
  \atlas{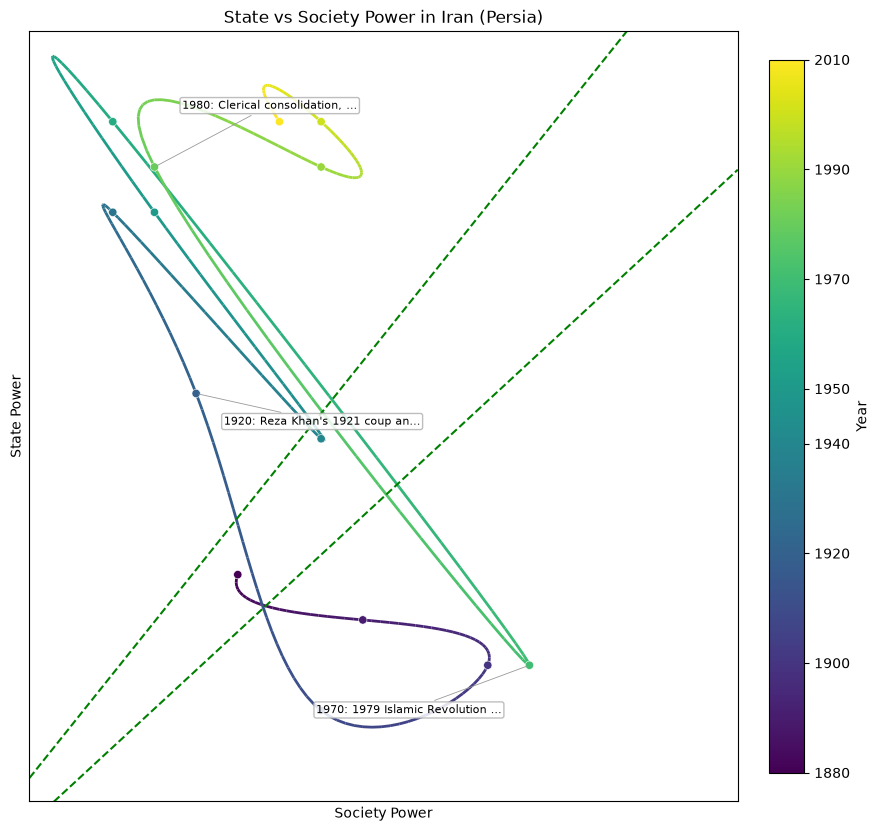}\hfill
  \atlas{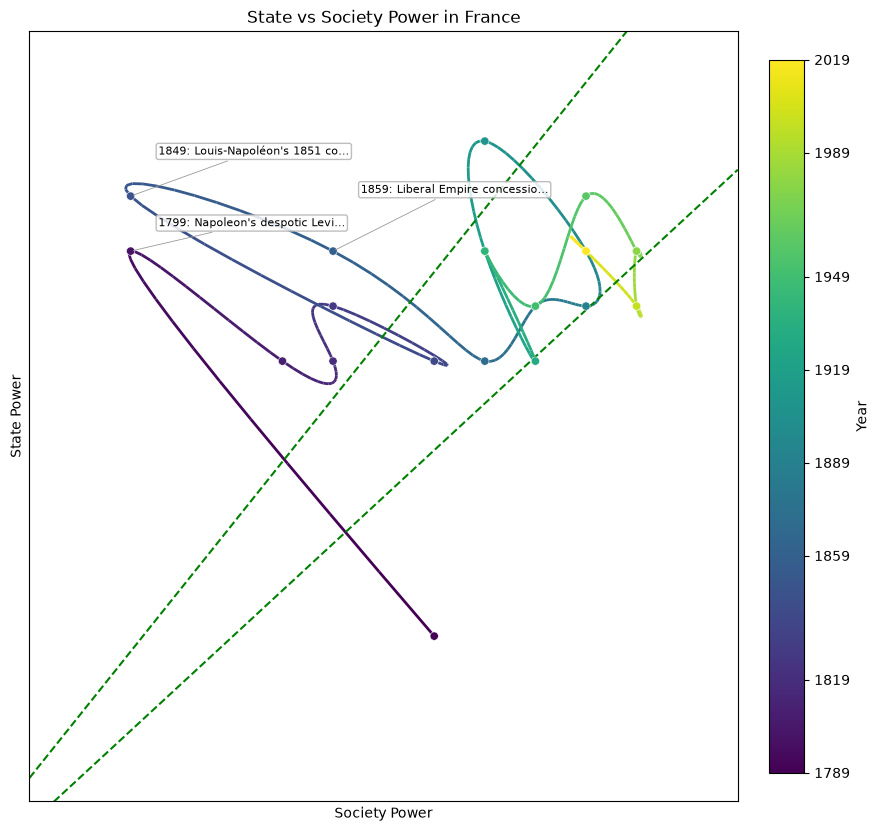}\hfill
  \atlas{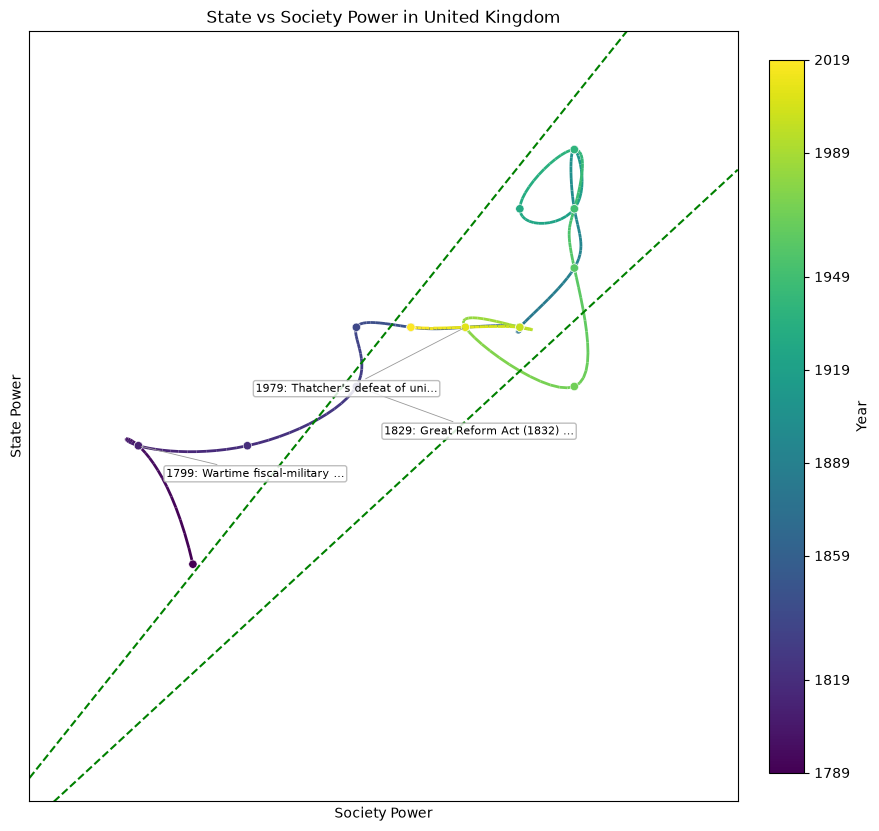}\\[6pt]
  \atlas{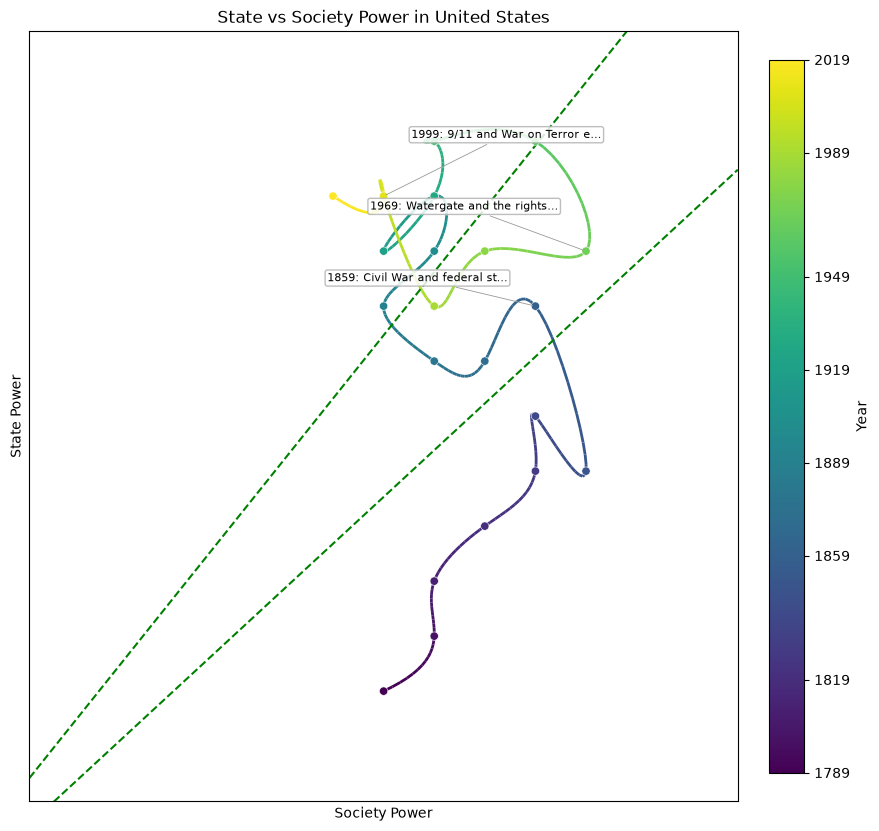}\hfill
  \atlas{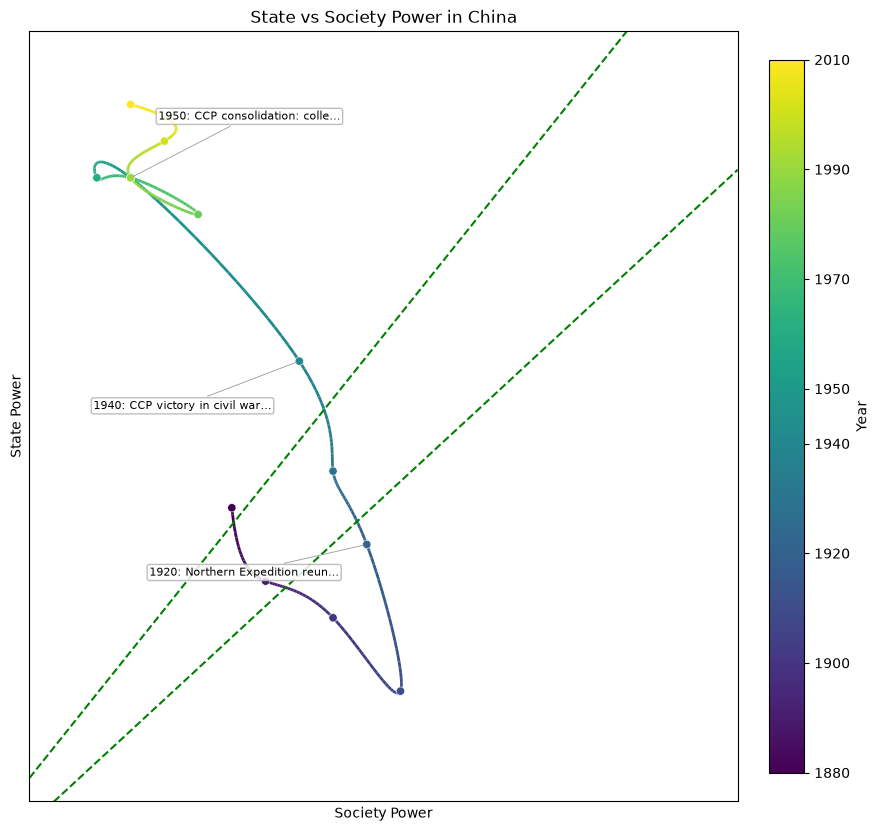}\hfill
  \atlas{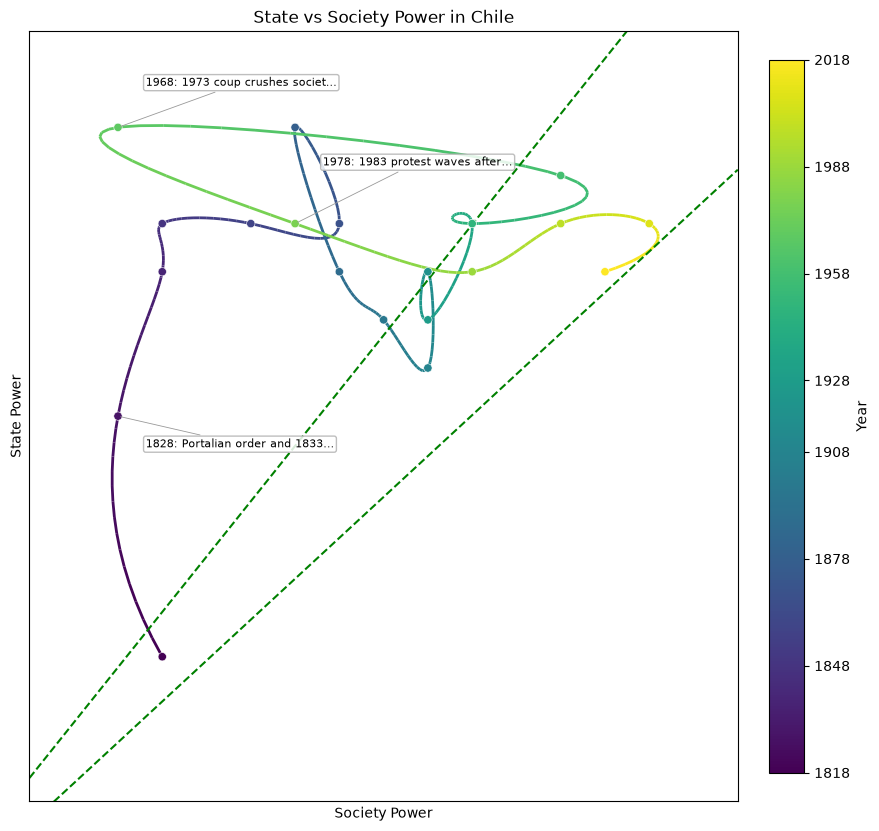}\\[6pt]
  \atlas{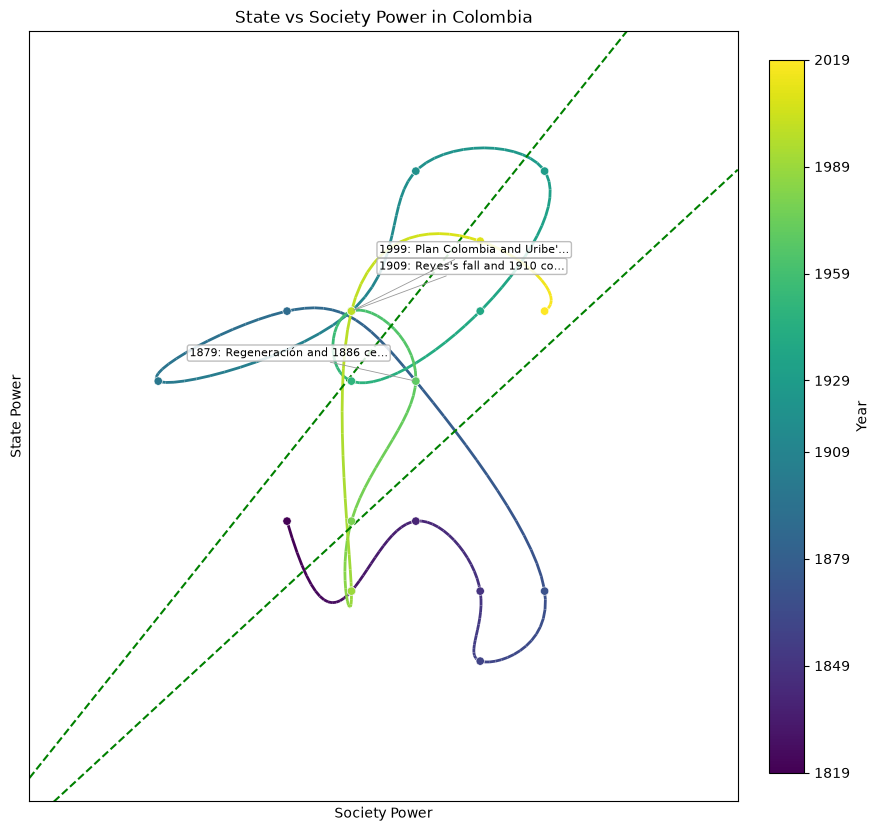}\hfill
  \atlas{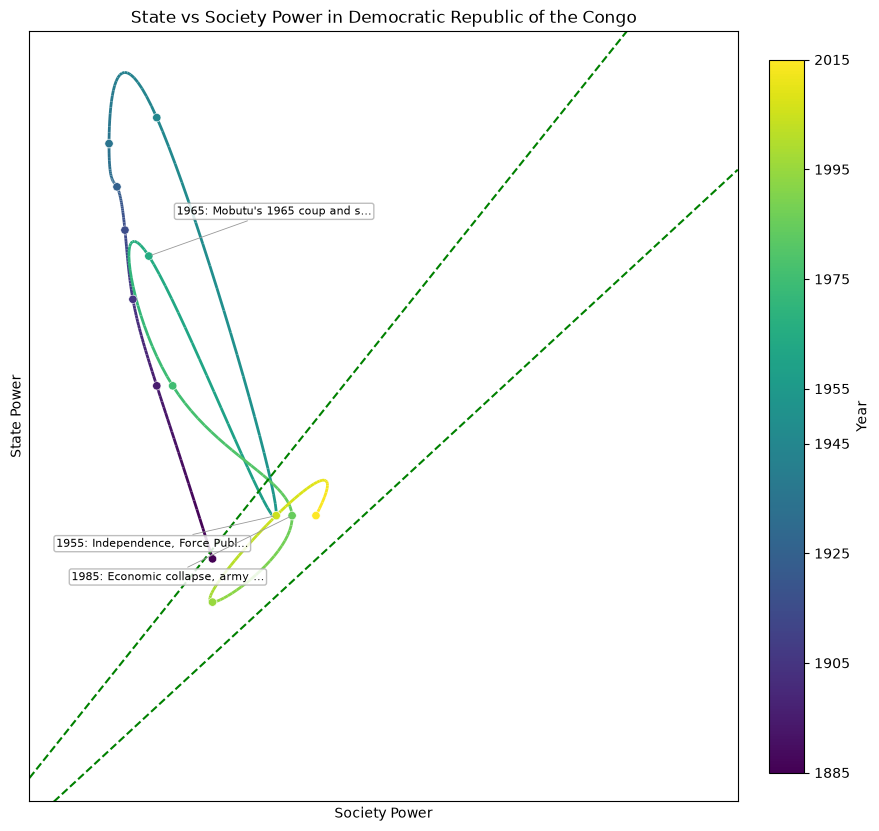}\hfill
  \atlas{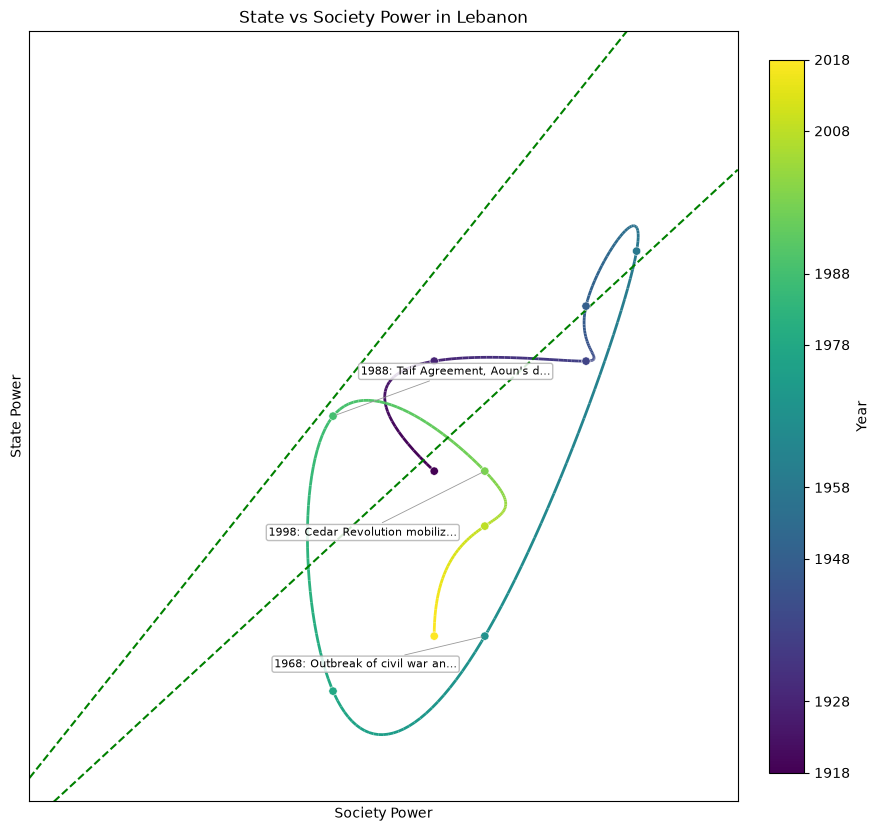}\\[6pt]
  \atlas{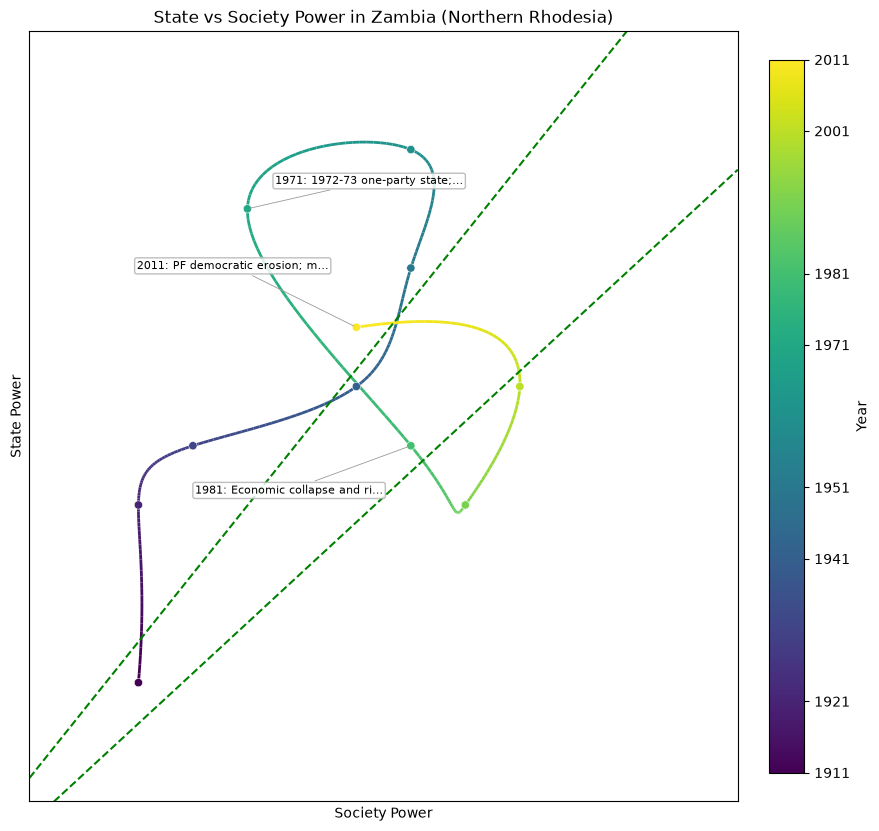}\hfill
  \atlas{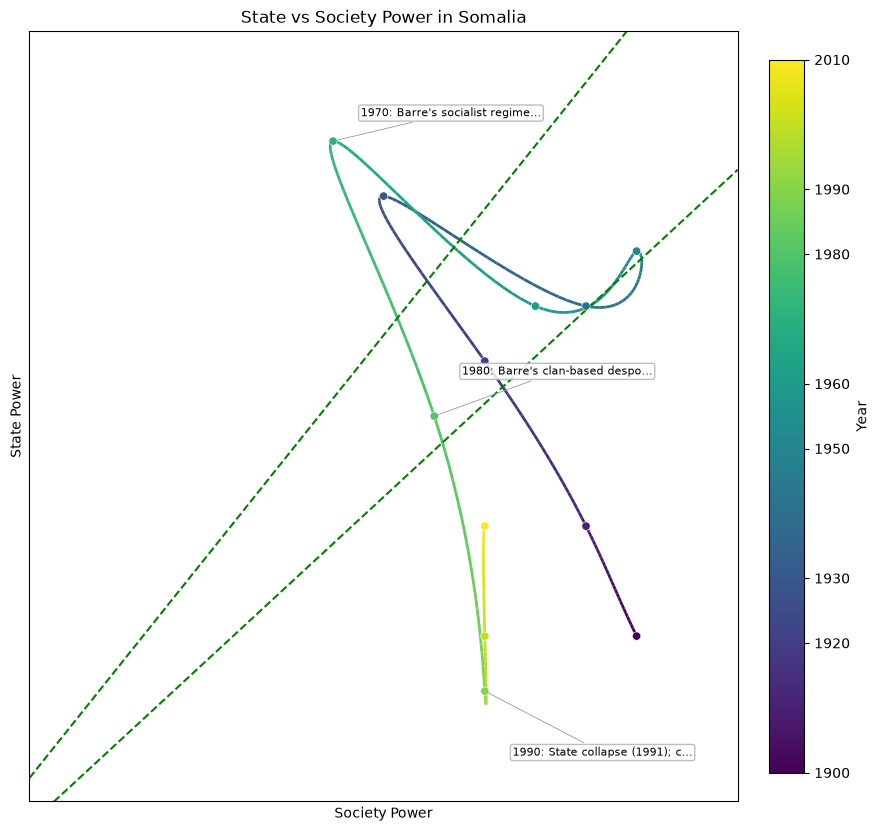}\hfill
  \atlas{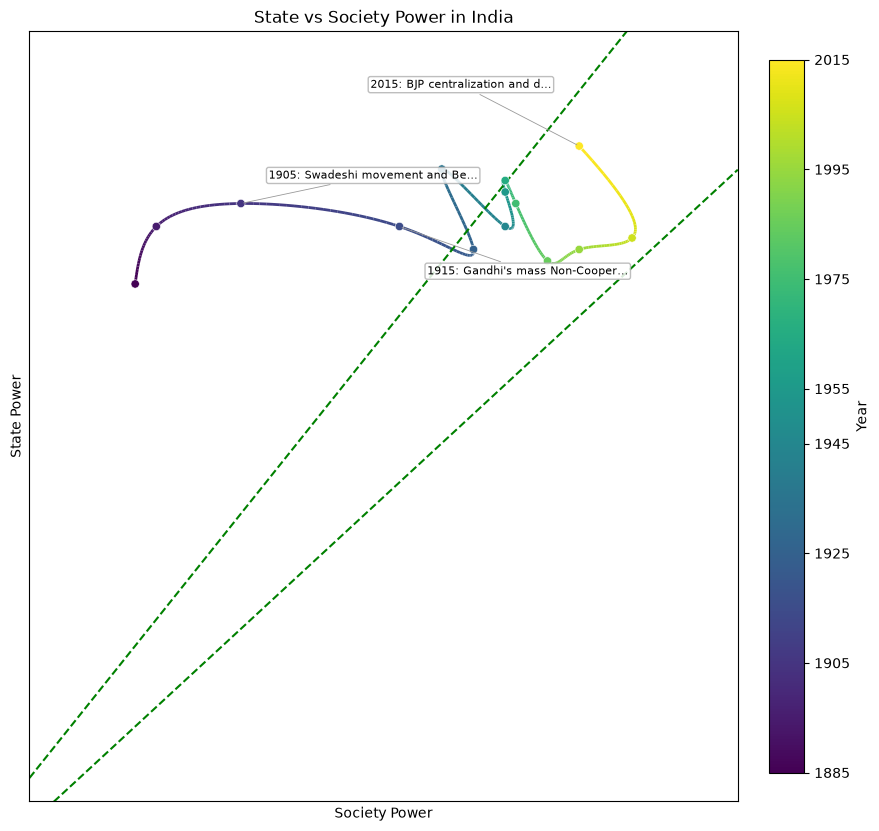}
  \caption{Trajectory atlas (reference model \texttt{claude-opus-4-8}): society
  power (x) vs.\ state power (y) for all twelve countries, time-shaded (dark $=$
  early, bright $=$ late); dashed lines mark the book's widening corridor
  illustratively. Animated versions are in the online gallery.}
  \label{fig:atlas}
\end{figure*}

Figure~\ref{fig:atlas} shows the twelve trajectories under \texttt{claude-opus-4-8},
chosen as a single reference model for legibility; every model's per-country
panels are in the online gallery, with cross-model disagreement quantified in
Section~\ref{sec:reliability}. Read against the book's four Leviathan types, most
placements land where the framework expects. The \textbf{United Kingdom} stays
inside the corridor throughout, state and society rising together (the Red
Queen), a textbook Shackled Leviathan. \textbf{France} enters it, moving from
society-leaning in 1789 to inside by the late twentieth century, Shackled by its
late history. \textbf{Iran} and \textbf{China} exit \emph{upward} into state
dominance, Iran after 1979 and China after 1949, the Despotic Leviathan the book
assigns them. \textbf{Chile} loops out and back, a sharp fall of society and
spike of state at the 1973 coup, then a return toward the corridor after the 1990
transition. \textbf{Colombia}, the book's Paper Leviathan, rises from a weak post-independence
founding to only moderate levels by 2019 (society $\approx6.4$, state
$\approx5.7$, the 1886 centralist constitution its largest move), never
consolidating the strong, high-capacity state of the Shackled cases; it ends in
the lower corridor rather than climbing to the strong-both corner, the muted arc
the Paper-Leviathan reading predicts, though the model treats late-twentieth-century
Colombia as closer to the corridor than clearly below it. The \textbf{Democratic
Republic of the Congo}, the book's Absent Leviathan, stays near the origin
throughout: state power spikes briefly under Mobutu's mid-century regime, then
collapses around 1960 independence and the subsequent wars, ending low on both
axes (society $\approx3.2$, state $\approx3.2$), the persistently weak state the
Absent-Leviathan label describes. The \textbf{United States} is another
Shackled Leviathan, much like the UK: state and society grow strong together and
it ends inside the corridor, if with a somewhat stronger state tilt than the UK.
Its state axis climbs faster than the UK's, but it stays within the corridor
rather than crossing into state dominance. \textbf{Lebanon} and \textbf{Somalia}
end deep in the weak-state region below the corridor (society $\approx4.8$ and
$4.6$, state $\approx1.9$ and $2.6$), a society outrunning a state that is barely
present: Lebanon's state erodes across the 1975--90 civil war and the post-2019
collapse, Somalia's stays low and flat throughout. They illustrate the
Paper/Absent point directly, sitting in the same low-state region while differing
only \emph{de jure} (Section~\ref{sec:setup}). \textbf{Zambia} is the case that
may surprise readers unfamiliar with its recent history: hypothesized a Paper
Leviathan, the models instead carry it \emph{into} the corridor by 2020 (society
$\approx5.8$, state $\approx6.0$), reading its post-1991 multiparty democracy and
its peaceful transfers of power as genuine balanced growth rather than a hollow
paper state, a reading that is defensible on that history even though it does not
match the label we assigned in advance. \textbf{India} lands where the book
places it, a Shackled Leviathan (society $\approx8.0$, state $\approx8.5$): a
strong colonial state in 1885 that its society gradually catches and checks,
ending high on both axes. Long diagonal runs mark co-growth; tight loops mark
oscillation in contested periods; large single-period jumps mark ruptures.

\subsection{Ensemble atlas (mean across models)}
\label{sec:ensemble}
\begin{figure*}[p]
  \centering
  \atlas{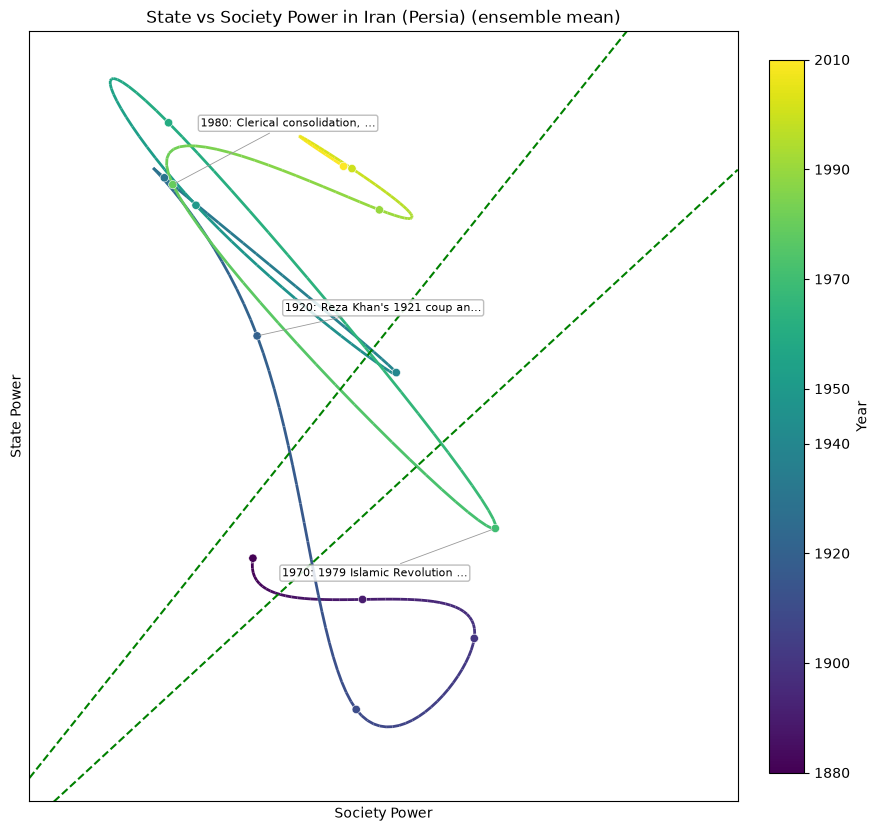}\hfill
  \atlas{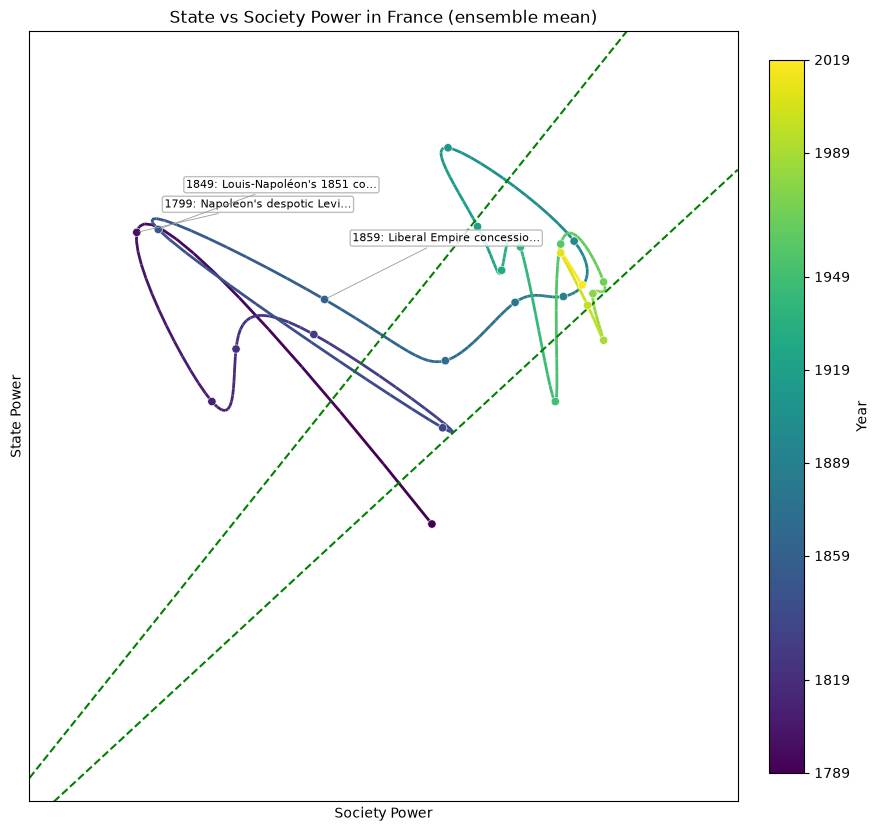}\hfill
  \atlas{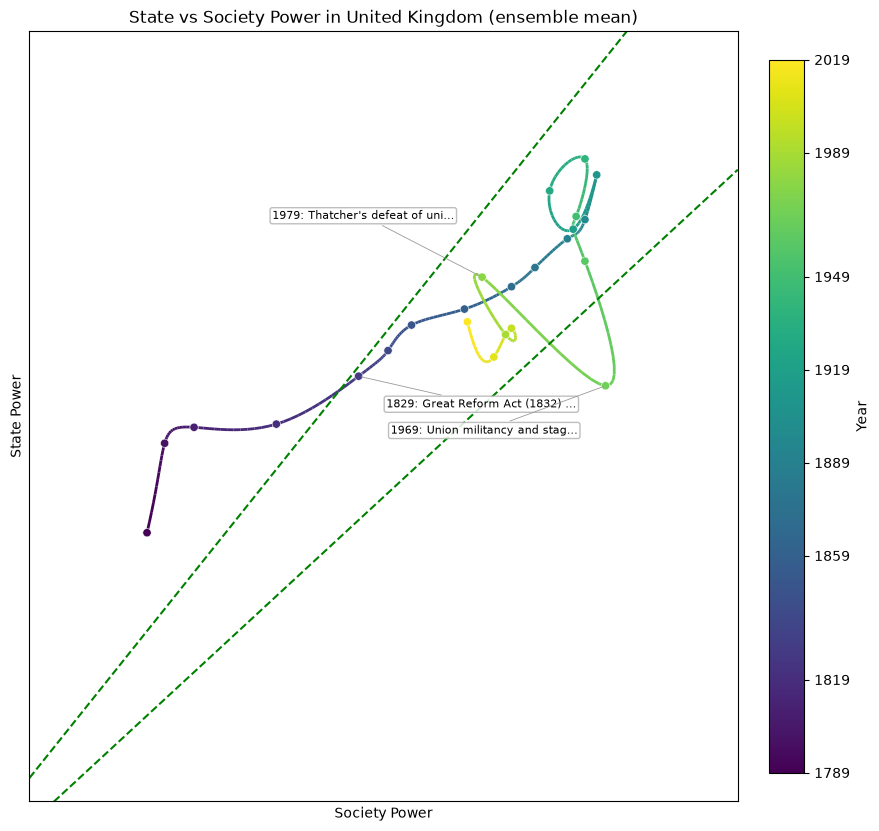}\\[6pt]
  \atlas{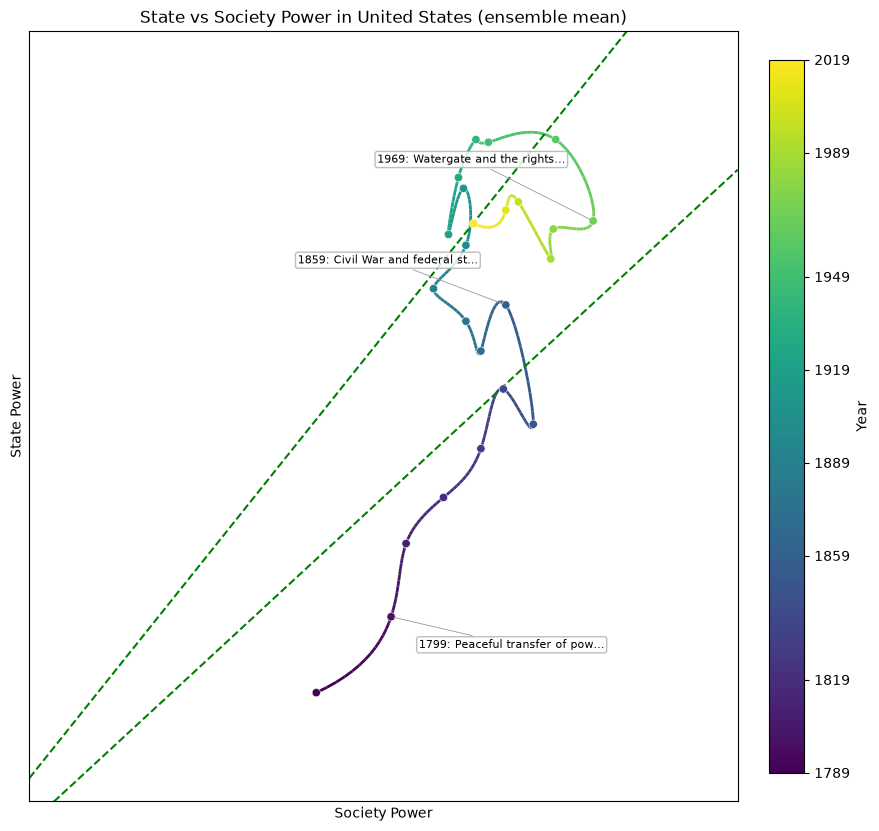}\hfill
  \atlas{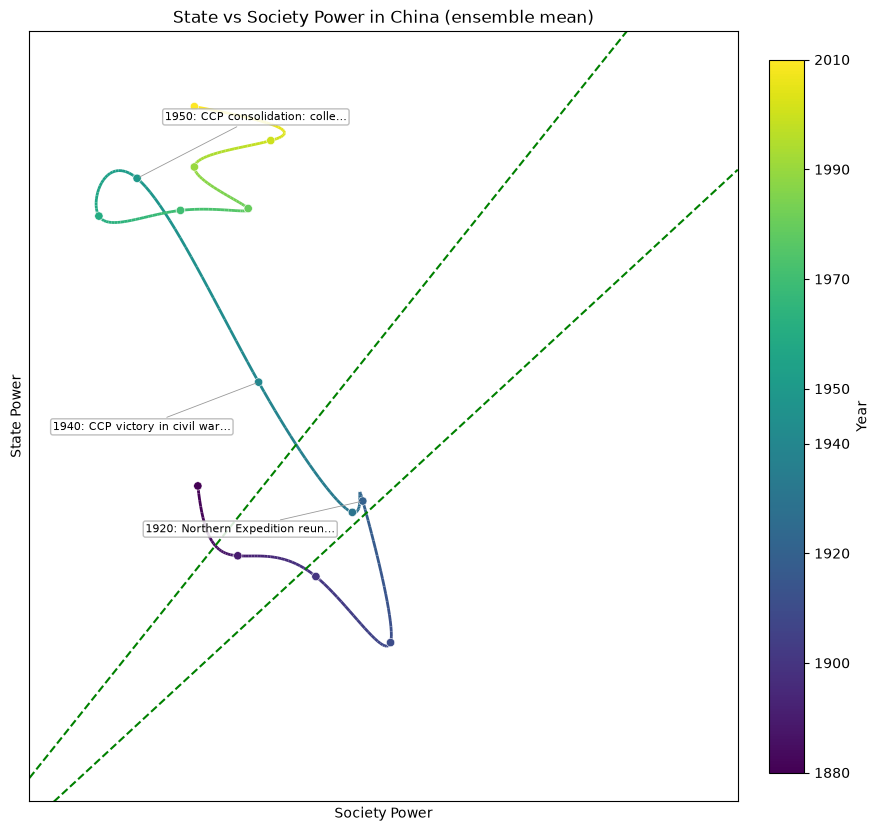}\hfill
  \atlas{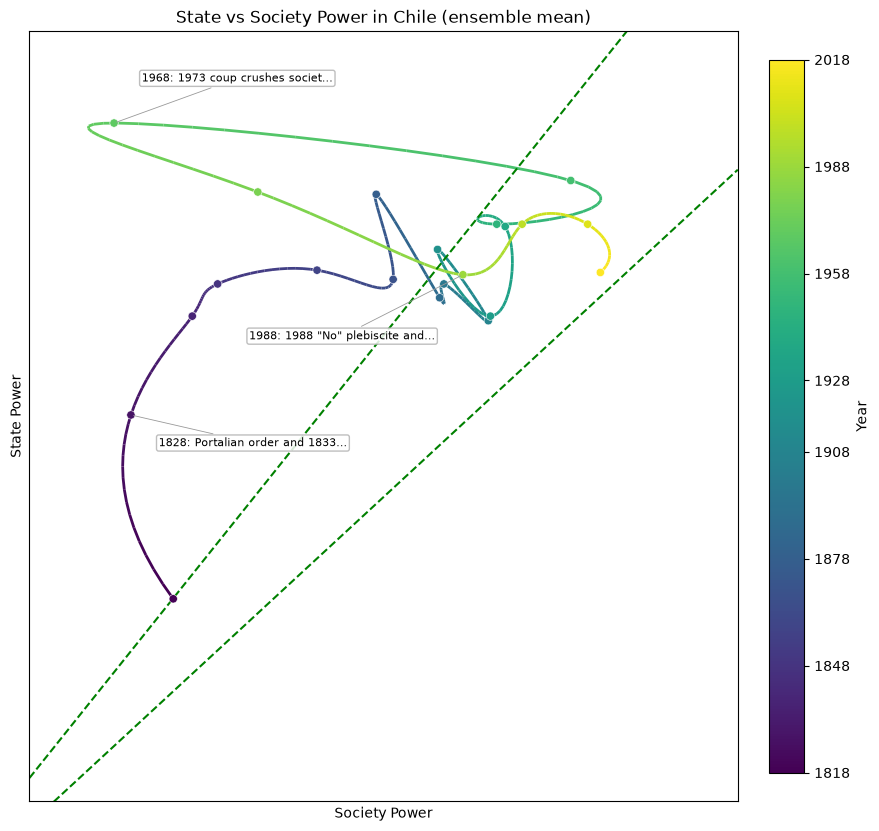}\\[6pt]
  \atlas{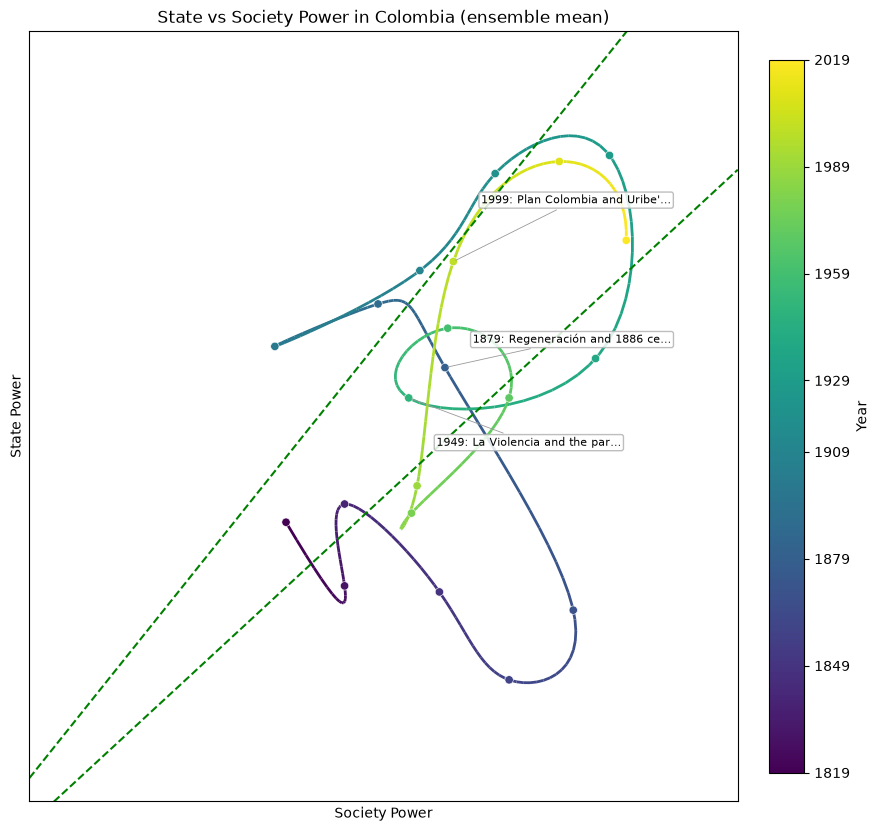}\hfill
  \atlas{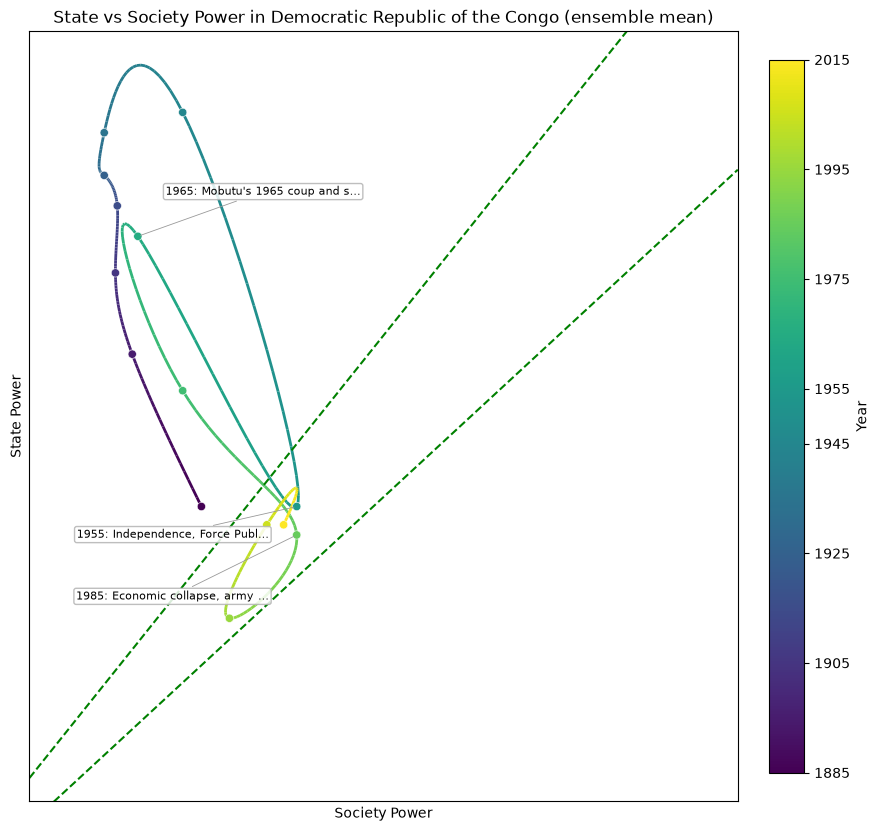}\hfill
  \atlas{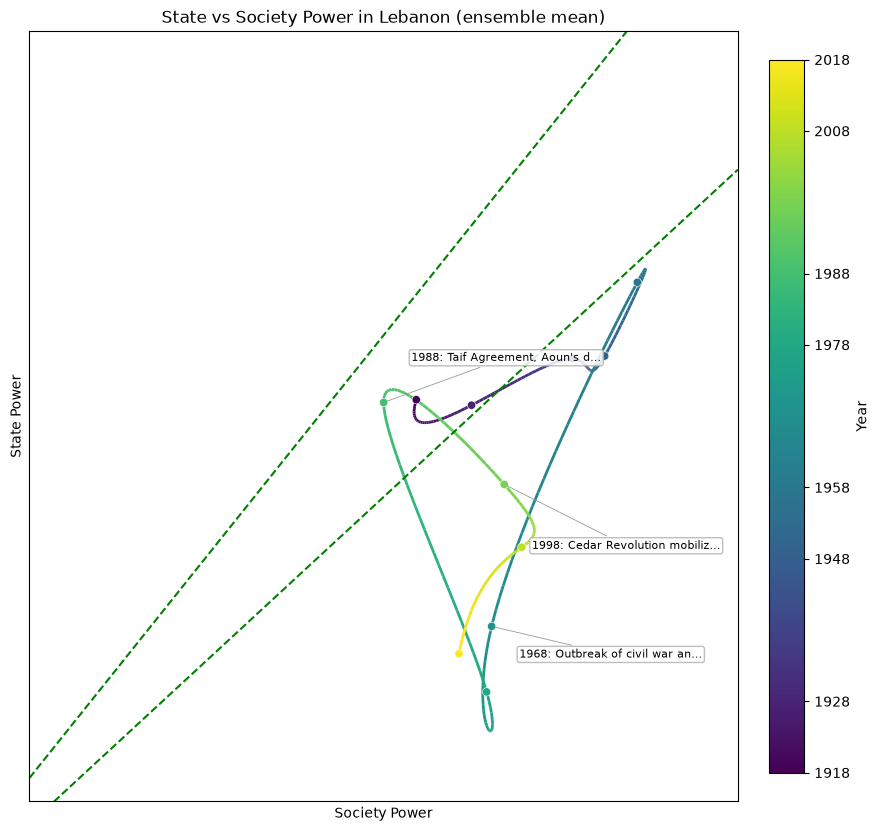}\\[6pt]
  \atlas{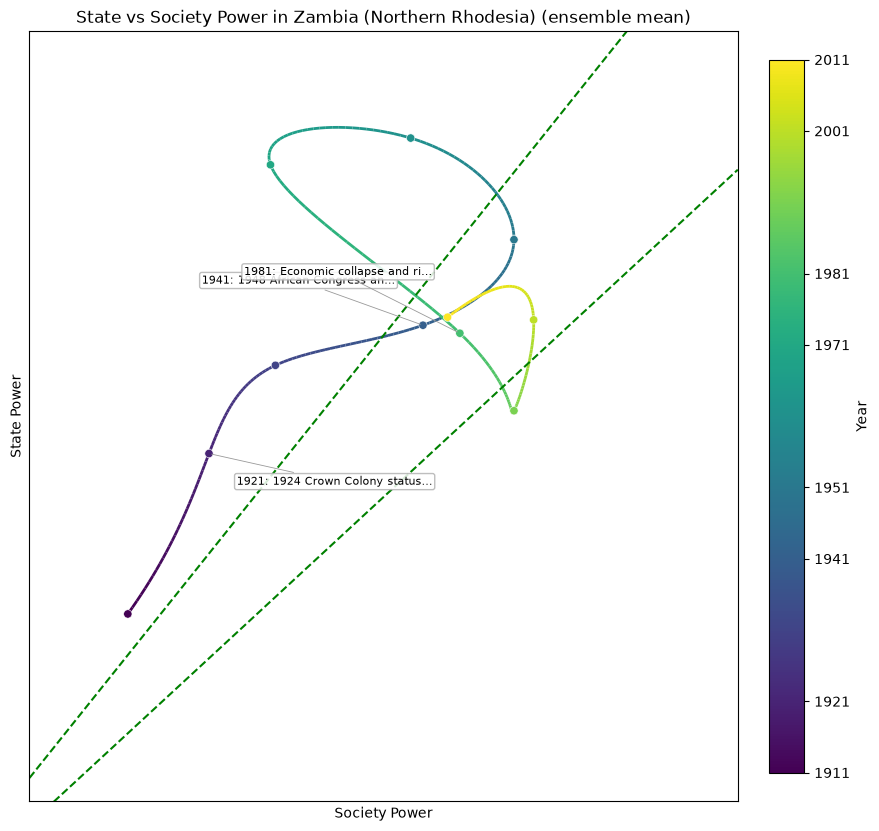}\hfill
  \atlas{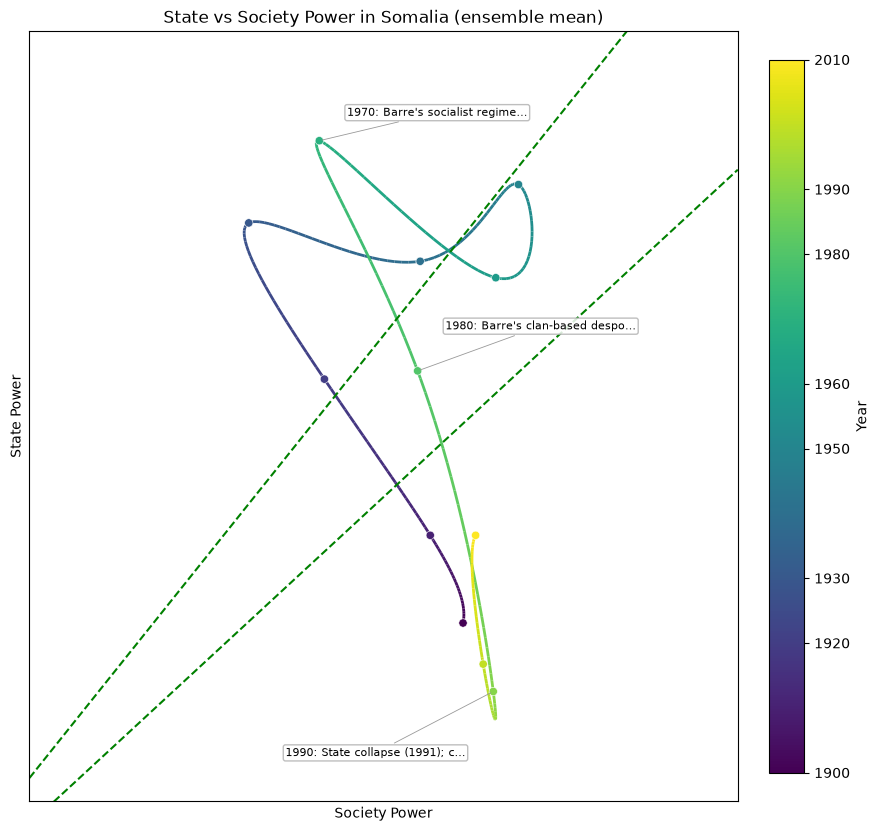}\hfill
  \atlas{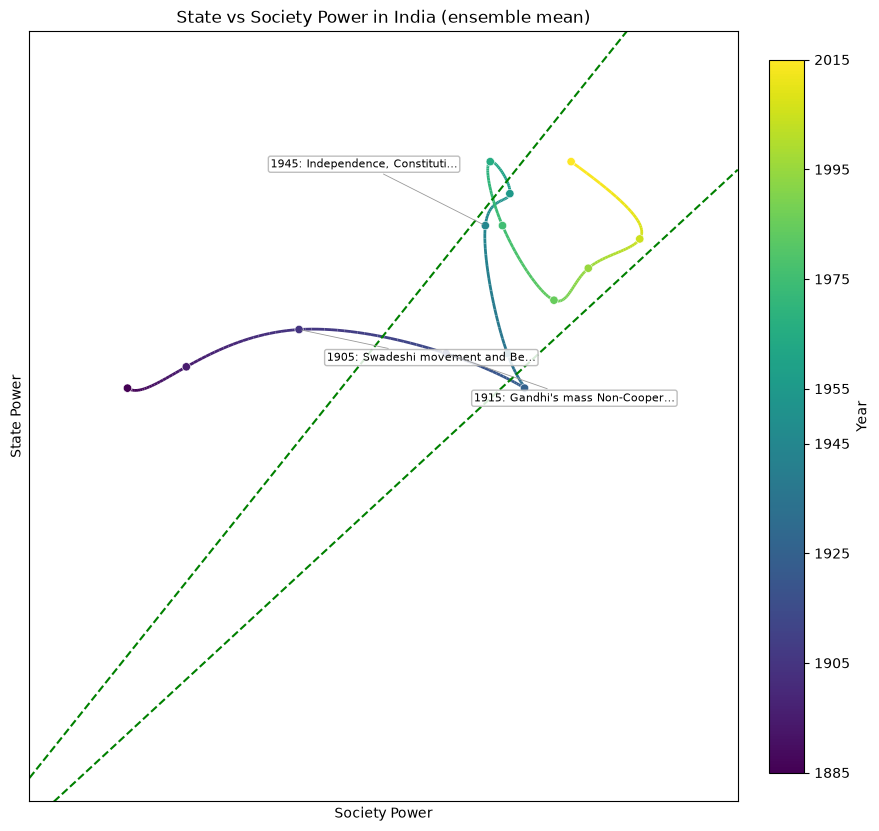}
  \caption{Ensemble atlas: the same twelve countries, each averaged across the four
  models period by period (panels titled ``ensemble mean''). The mean damps any
  one model's idiosyncrasy; the spread it averages over is Krippendorff's $\alpha$
  in Table~\ref{tab:intermodel}. Time-shaded; dashed lines mark the widening
  corridor.}
  \label{fig:atlas-mean}
\end{figure*}

Figure~\ref{fig:atlas-mean} shows, for each country, the trajectory averaged
across all four models period by period. Because independent models rarely share
the same idiosyncratic error, the mean is a more robust reading than any single
model's path: where the four agree it tracks the consensus tightly, and where
they diverge it sits between them, so a point should then be read as uncertain
rather than precise (the underlying per-axis agreement is quantified as
Krippendorff's $\alpha$ in Section~\ref{sec:reliability}). Each panel is titled
``(ensemble mean)'' to distinguish it from the single-model atlas of
Figure~\ref{fig:atlas}. We treat this mean
atlas, not any single model, as the paper's headline reading. The mean is close
to the reference model for the high-agreement cases (Iran, Chile;
Section~\ref{sec:reliability}) and visibly smoother for the United States, where
the models disagree most about how fast its state grew. The corridor placements
are unchanged: the UK, France, and the US inside the corridor, Chile entering it,
Iran and China state-dominated, and Colombia and the Congo low.

\subsection{One map, all countries}
\label{sec:combined}
\begin{figure*}[t]
  \centering
  \IfFileExists{experiments/results/all-countries__ensemble-mean.png}%
    {\includegraphics[width=0.72\textwidth]{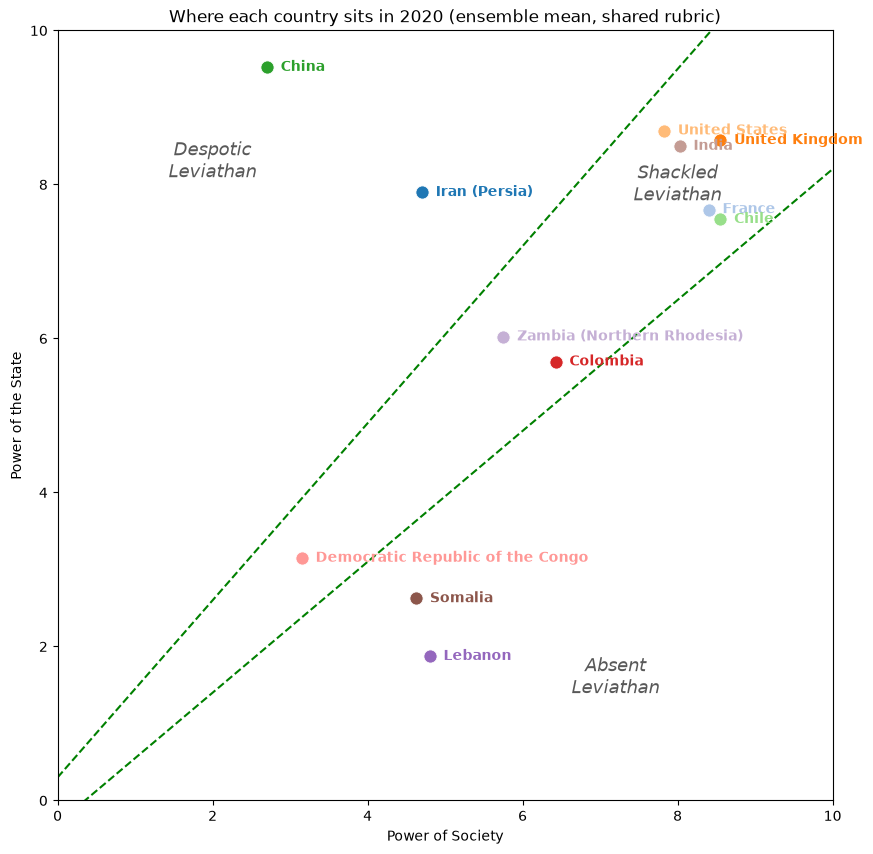}}%
    {\fbox{\parbox[c][6cm][c]{0.6\textwidth}{\centering\footnotesize
       \textcolor{red}{\texttt{all-countries\_\_ensemble-mean}}\\[2pt]
       (run \texttt{atlas\_combined.py})}}}%
  \caption{All twelve countries' ensemble-mean trajectories in one shared
  (society, state) space, the direct analogue of the book's single-plane map. An
  open marker is each country's start year, a filled marker and label its last
  period; dashed lines mark the book's widening corridor. Because every country is
  scored against the same fixed rubric, the paths are plotted on a common scale
  without any per-country rescaling.}
  \label{fig:combined}
\end{figure*}

The book plots countries together in a single plane. Doing the same requires the
scores to be comparable \emph{across} countries, not merely across periods within
one country, and this is a real question rather than a given: does the shared
rubric actually put every country on one scale? Our design bets that it does. The
fixed $0$--$10$ anchors (Section~\ref{sec:method}) are country-independent, so a
$7$ on the state axis is meant to denote the same level of state capacity for the
Congo as for China. Figure~\ref{fig:combined} overlays all twelve ensemble-mean
paths in one space to test that bet. The relative placement is broadly sensible:
the Despotic cases (China, Iran) sit high on the state axis with weak society, the
Shackled cases (UK, France, the US, and India) climb the diagonal into the
strong-both upper-right, and the weak-state cases fall below and left of the
corridor, the Congo, Somalia, and Lebanon down in the low-state region (society
ahead of a barely-present state) and Colombia at the corridor's lower edge.
The one country the models place outside its candidate region is Zambia, inside
the corridor rather than below it as a Paper Leviathan, a reading that fits its
stable post-1991 democracy. That the countries otherwise sort
into the book's regions on a single common scale is evidence the rubric transports
across countries well enough for a joint reading. It is not proof: the shared
training-data biases that may inflate a score could do so consistently across
countries (Section~\ref{sec:limitations}), and the absolute levels are softer than
the within-country ordering. We therefore read the joint map as a qualitative
cross-country atlas, not a metric one.

Three regularities recur. First, the countries the book treats as
corridor-dwellers (the UK throughout, France and Chile by their late histories)
trace long diagonal runs in which state and society grow together. Second,
every country's largest single-period move is an institutional rupture that
pushes the \emph{state} axis up: the 1979 Iranian revolution, the 1973 Chilean
coup, post-1949 Chinese consolidation, and the US Civil War. Third, the two
twentieth-century revolutions (Iran 1979, China 1949) end in state dominance
rather than liberty. The mechanism the paths trace is not that mobilizing society
produces despotism, which would invert the book's logic, but the opposite: a
revolution briefly raises society's power, then the new revolutionary state
consolidates and \emph{demobilizes} the society that carried it, outgrowing
rather than balancing it. Liberty fails because society does not stay strong
enough to shackle the state the revolution builds, which is the book's Despotic
Leviathan path.

\subsection{Consistency with V-Dem}
\label{sec:validation}
\begin{figure*}[p]
  \centering
  \atlas{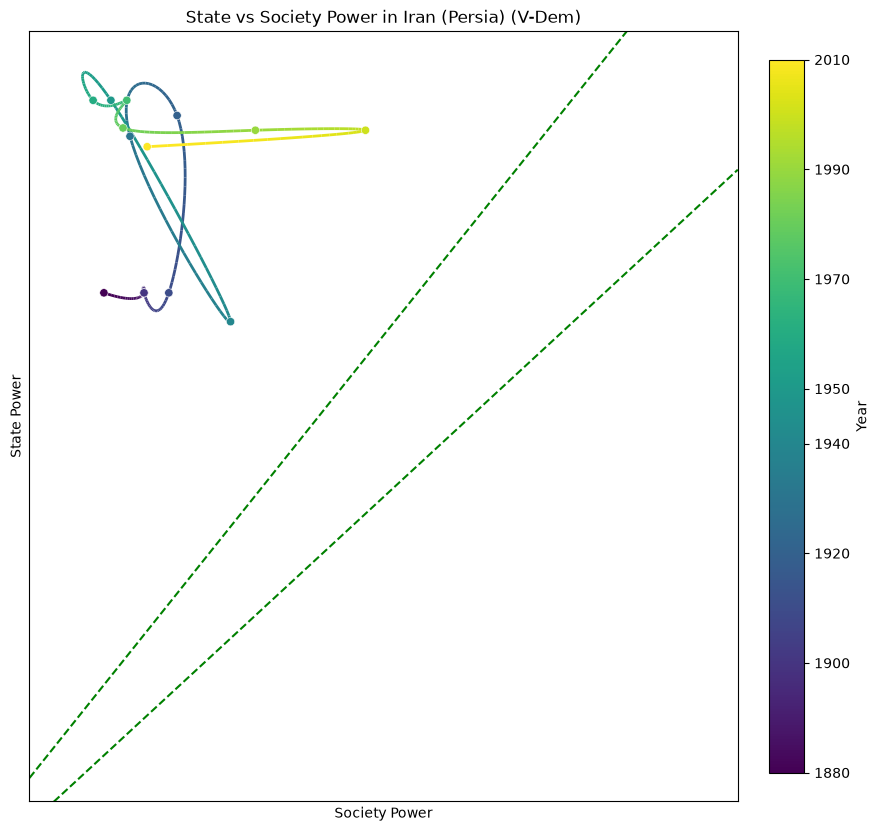}\hfill
  \atlas{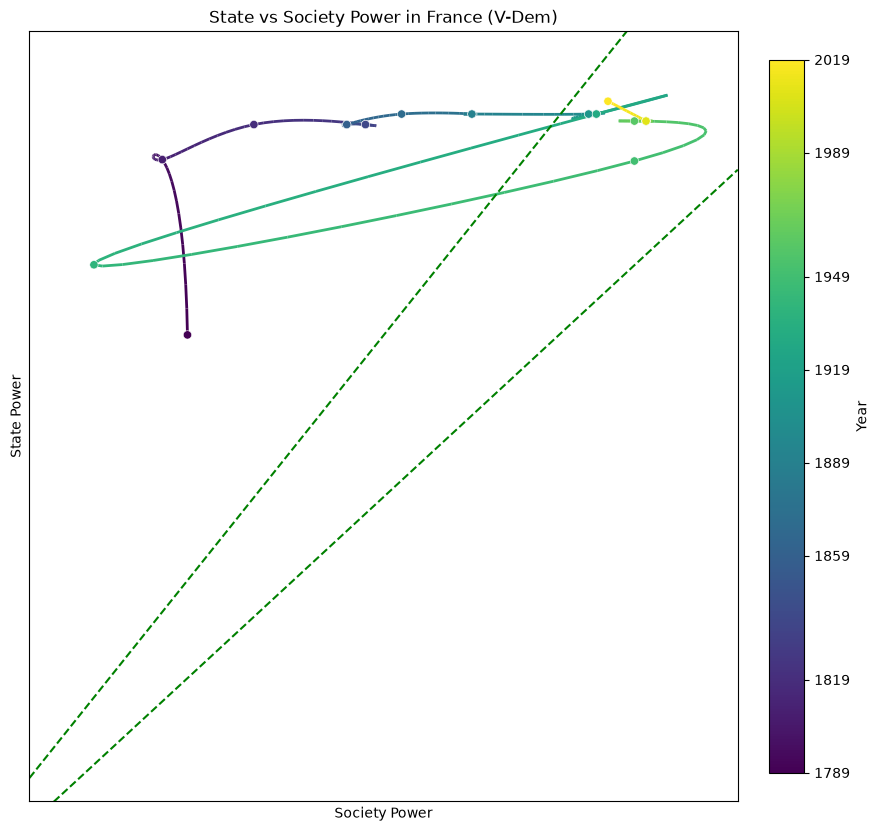}\hfill
  \atlas{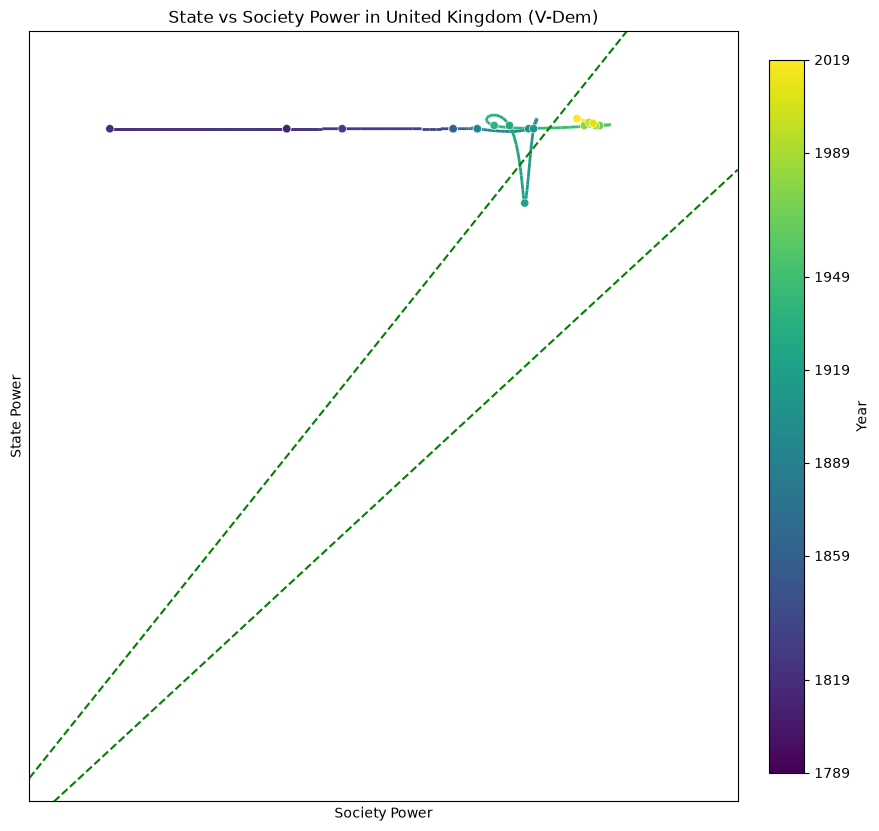}\\[6pt]
  \atlas{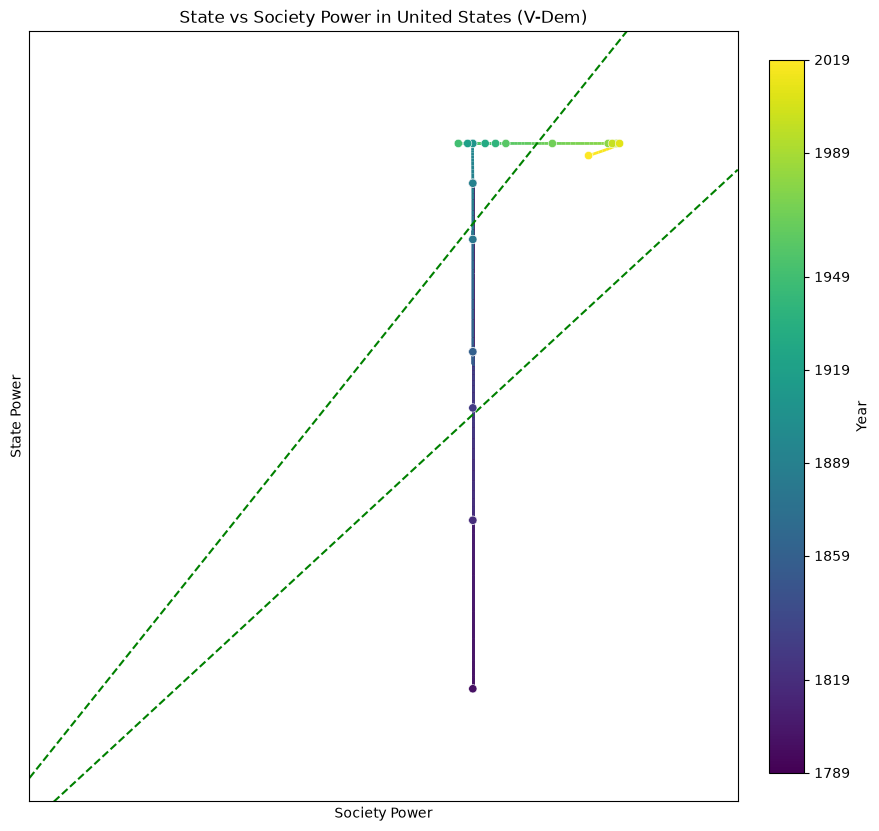}\hfill
  \atlas{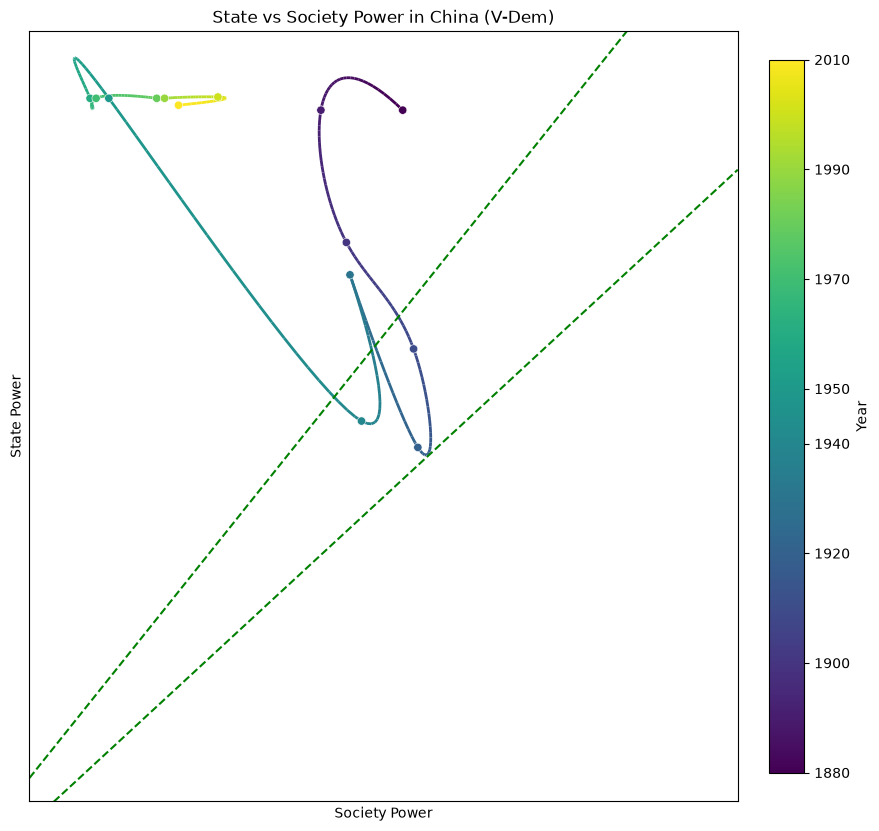}\hfill
  \atlas{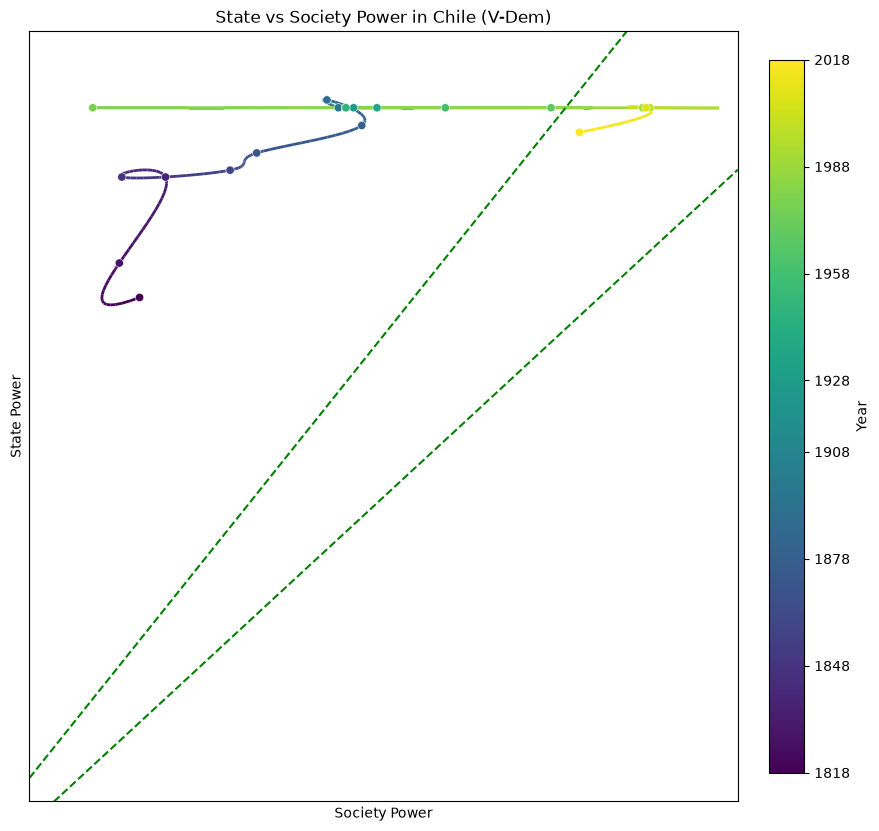}\\[6pt]
  \atlas{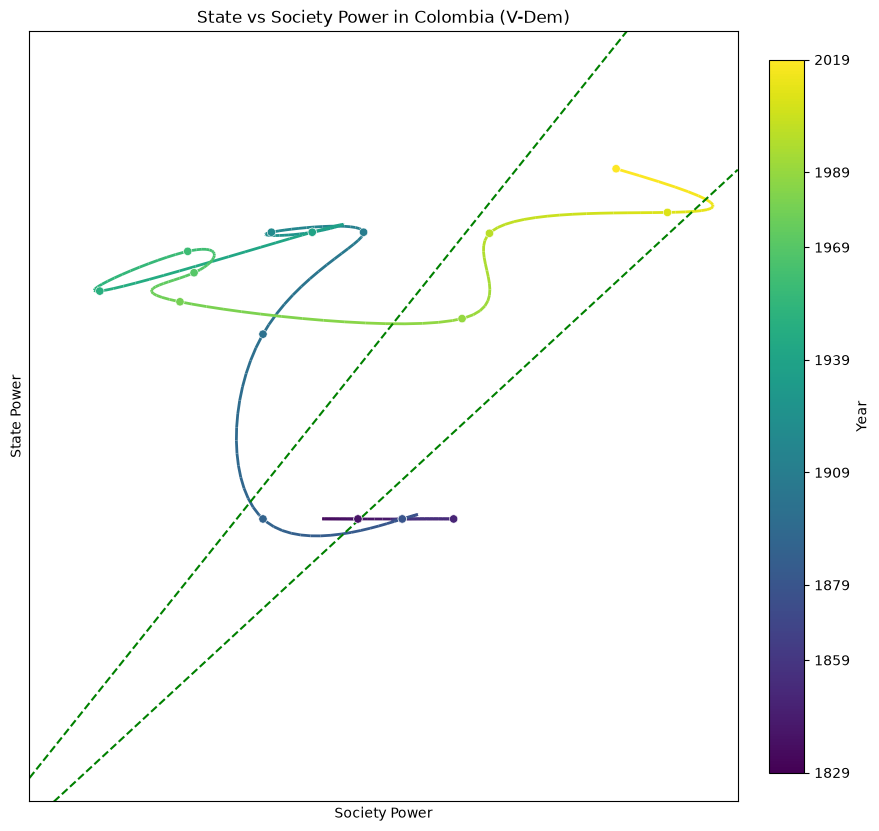}\hfill
  \atlas{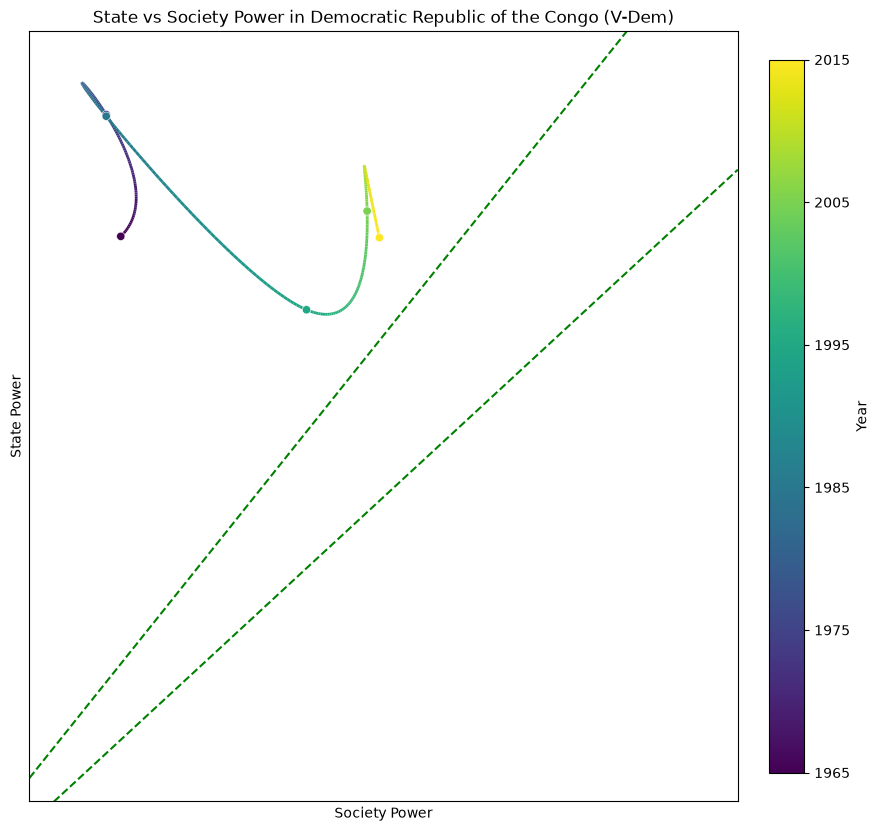}\hfill
  \atlas{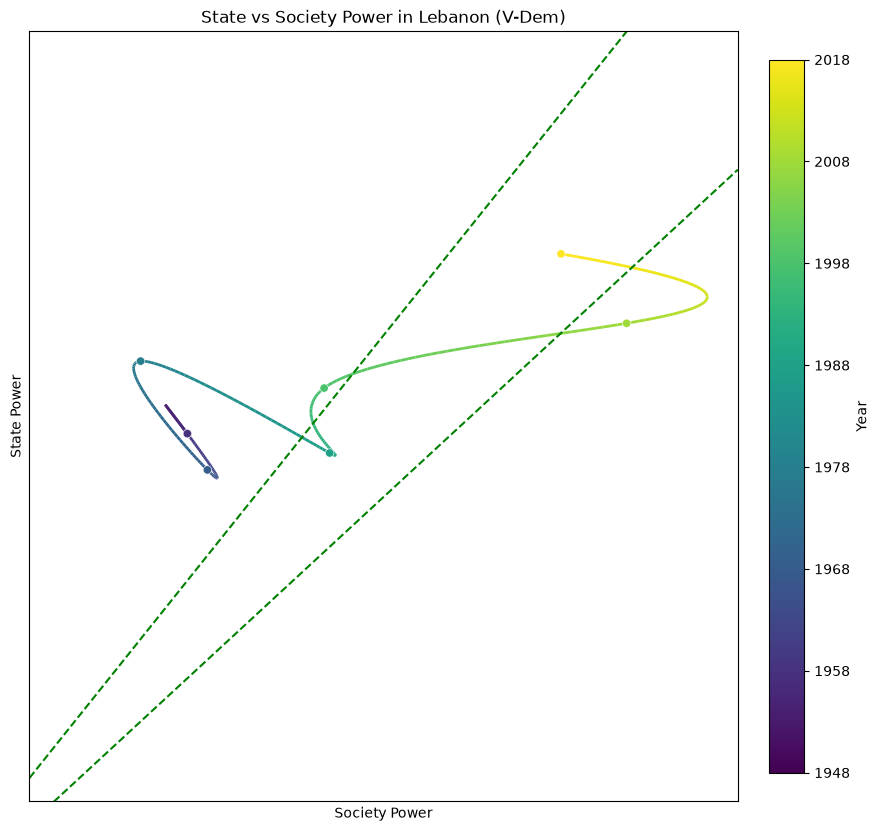}\\[6pt]
  \atlas{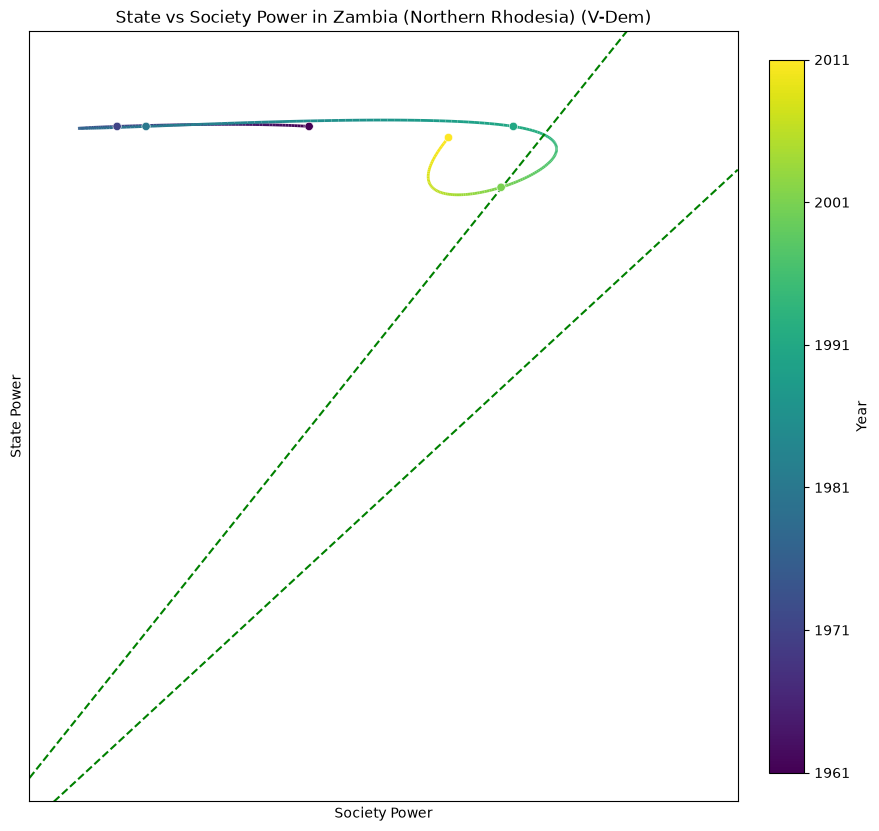}\hfill
  \atlas{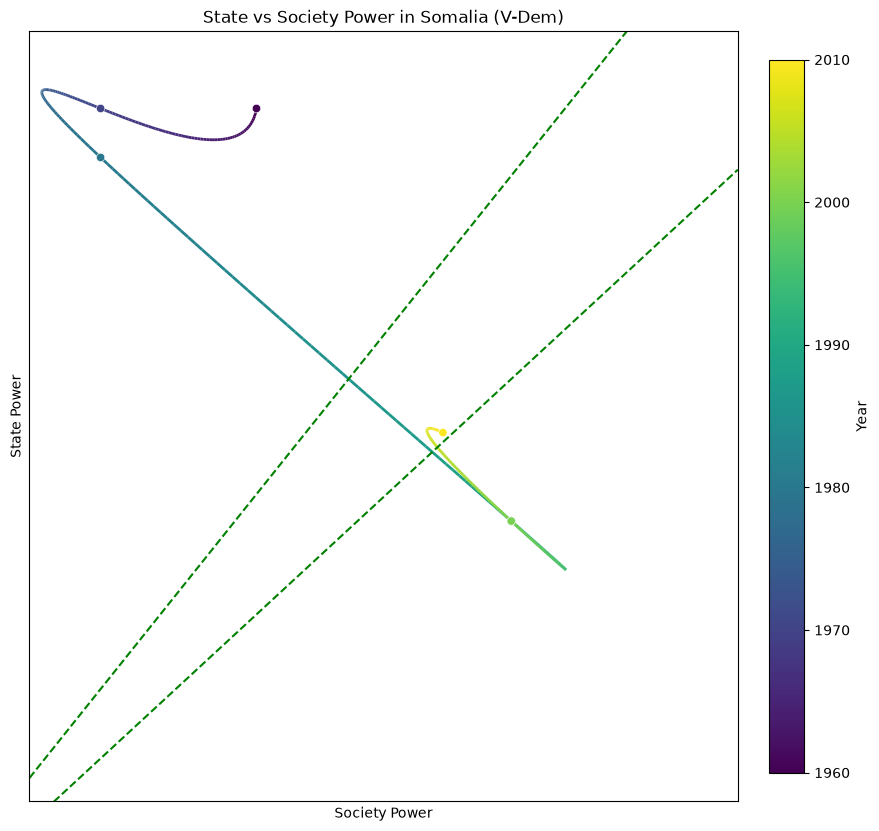}\hfill
  \atlas{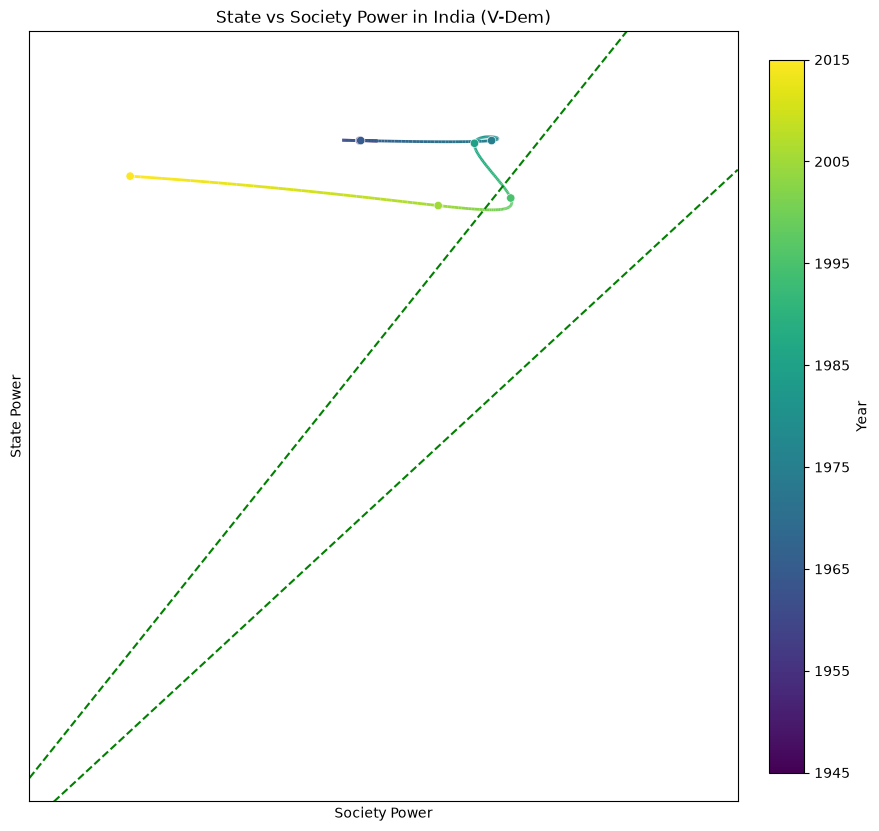}
  \caption{V-Dem atlas: the same twelve countries from expert codings (society $=$
  core civil-society index $\times10$; state $=$ authority over territory $/10$)
  at the LLM period-midpoint years. The society axis broadly agrees with the LLM
  ensemble (Fig.~\ref{fig:atlas-mean}), but V-Dem's territorial-authority state
  axis is near-saturated for consolidated states and misses the state-capacity
  dynamics the LLM captures.}
  \label{fig:vdem}
\end{figure*}

Table~\ref{tab:validation} reports correlations between the LLM scores and V-Dem
over overlapping years, on both levels and first differences. On \emph{levels},
LLM society power tracks V-Dem's core civil-society index strongly for China
($\rho=0.89$), France ($0.87$), Chile ($0.74$) and the UK ($0.70$); the one low
value, the United States ($\rho=-0.10$), is a variance artifact rather than a
category mismatch, since V-Dem holds US civil society nearly flat and high across
the whole window, leaving almost nothing for a rank correlation to track. On
the more demanding \emph{first differences} the society signal is positive but
weaker (China $0.66$, Iran $0.60$, France $0.40$), as expected once the shared
trend is removed. The \emph{state} axis is where consistency collapses:
first-difference $\rho\approx0$ for most countries. Figure~\ref{fig:vdem}
shows why: our V-Dem state proxy (\texttt{v2svstterr}, authority over territory)
is near-saturated and flat for consolidated states, so it cannot move with the
LLM's broader state-\emph{capacity} notion; the LLM tracks the \emph{building} of
the modern state that a territorial-control index does not. As noted above, this
measures agreement with expert consensus the models were likely trained on, not
independent accuracy.
\begin{table*}[t]
\centering
\small
\caption{Consistency with V-Dem (\texttt{claude-opus-4-8}): Spearman $\rho$ at period-midpoint years, reported as level\,/\,$\Delta$ (levels vs.\ first differences). Society vs.\ \texttt{v2xcs\_ccsi}; state vs.\ \texttt{v2svstterr}. First differences are the more honest signal.}
\label{tab:validation}
\begin{tabular}{lcccc}
\toprule
Country & Window & $N$ & Society $\rho$ & State $\rho$ \\
\midrule
Iran (Persia) & 1880--2020 & 14 & 0.45\,/\,0.60 & 0.42\,/\,0.13 \\
France & 1789--2020 & 24 & 0.87\,/\,0.40 & 0.20\,/\,0.04 \\
United Kingdom & 1789--2020 & 24 & 0.70\,/\,0.21 & 0.13\,/\,0.07 \\
United States & 1789--2020 & 24 & -0.10\,/\,0.26 & 0.84\,/\,0.15 \\
China & 1880--2020 & 14 & 0.89\,/\,0.66 & 0.74\,/\,0.24 \\
Chile & 1818--2020 & 21 & 0.74\,/\,0.17 & 0.18\,/\,0.42 \\
Colombia & 1819--2020 & 20 & 0.46\,/\,0.51 & 0.79\,/\,0.41 \\
DR Congo & 1885--2020 & 6 & 0.46\,/\,-0.30 & 0.52\,/\,-0.21 \\
Lebanon & 1918--2020 & 8 & -0.20\,/\,0.24 & -0.39\,/\,-0.33 \\
Zambia & 1911--2020 & 6 & 0.75\,/\,0.79 & 0.03\,/\,-0.23 \\
Somalia & 1900--2020 & 6 & 0.56\,/\,0.76 & 0.97\,/\,0.79 \\
India & 1885--2020 & 8 & 0.31\,/\,0.05 & 0.41\,/\,0.54 \\
\bottomrule
\end{tabular}
\end{table*}

Figure~\ref{fig:vdem} re-plots each country from V-Dem in the same space
(society $=$ \texttt{v2xcs\_ccsi}$\times10$, state $=$ \texttt{v2svstterr}$/10$),
at the LLM period-midpoint years. The society axis broadly rhymes with the LLM
ensemble (Fig.~\ref{fig:atlas-mean}), but the state axis is pinned high for every
consolidated state: V-Dem places even 1880s Qajar Iran and the early United
States near the top of the state axis, the opposite of the LLM, which reads those
as weak states still building capacity. The comparison is thus best read as
\emph{agreement on society, divergence on the operationalization of state power}.

\subsection{Model comparison and reliability}
\label{sec:reliability}
Treating each model as a ``rater'' in the sense of inter-coder
reliability~\citep{krippendorff2018content}, Table~\ref{tab:intermodel} reports
agreement between models on the same country. We report \emph{Krippendorff's}
$\alpha$, a standard inter-coder reliability coefficient that measures how much
raters (here, the four models) agree beyond what chance would give: $\alpha=1$ is
perfect agreement, $0$ is chance, and negative values mean systematic
disagreement, with $\alpha\ge0.8$ conventionally read as strong and
$0.67$--$0.8$ as tentative. We compute it (interval variant, treating the $0$--$10$
scores as interval-scaled) on the period-to-period \emph{changes} rather than on
levels (for the same reason first differences are used in
Section~\ref{sec:validation}): agreement on changes tests whether models see the
same \emph{moves}, not merely the same secular trend. Models agree
slightly more on society ($\alpha=0.70$ overall) than on the state axis
($0.68$), and most on the countries with the clearest arcs, Iran ($0.77/0.78$)
and Chile (society $0.83$).
Agreement is lowest for the United States (society $\alpha=0.47$), but the
disagreement is one of degree, not of category: all four models place it inside
the corridor as a Shackled Leviathan and differ only on how strong its central
and security state grew. The open-weight model does not systematically
diverge on the periods it scored.
\begin{table}[t]
\centering
\small
\caption{Inter-model reliability: Krippendorff's $\alpha$ (interval) on period-to-period \emph{changes} across models, per country. Higher $=$ models see the same moves.}
\label{tab:intermodel}
\begin{tabular}{lccc}
\toprule
Country & \#models & Society $\alpha$ & State $\alpha$ \\
\midrule
Iran (Persia) & 4 & 0.77 & 0.78 \\
France & 4 & 0.72 & 0.56 \\
United Kingdom & 4 & 0.56 & 0.63 \\
United States & 4 & 0.47 & 0.62 \\
China & 4 & 0.56 & 0.56 \\
Chile & 4 & 0.83 & 0.57 \\
Colombia & 4 & 0.54 & 0.53 \\
DR Congo & 4 & 0.51 & 0.77 \\
Lebanon & 4 & 0.34 & 0.71 \\
Zambia & 4 & 0.61 & 0.82 \\
Somalia & 4 & 0.56 & 0.83 \\
India & 4 & 0.47 & 0.16 \\
\midrule
\textbf{Overall} & -- & \textbf{0.66} & \textbf{0.68} \\
\bottomrule
\end{tabular}
\end{table}

\subsection{Prompt-language sensitivity}
\label{sec:language}
\begin{figure*}[t]
  \centering
  \IfFileExists{experiments/results_lang/lang-ablation.png}%
    {\includegraphics[width=\textwidth]{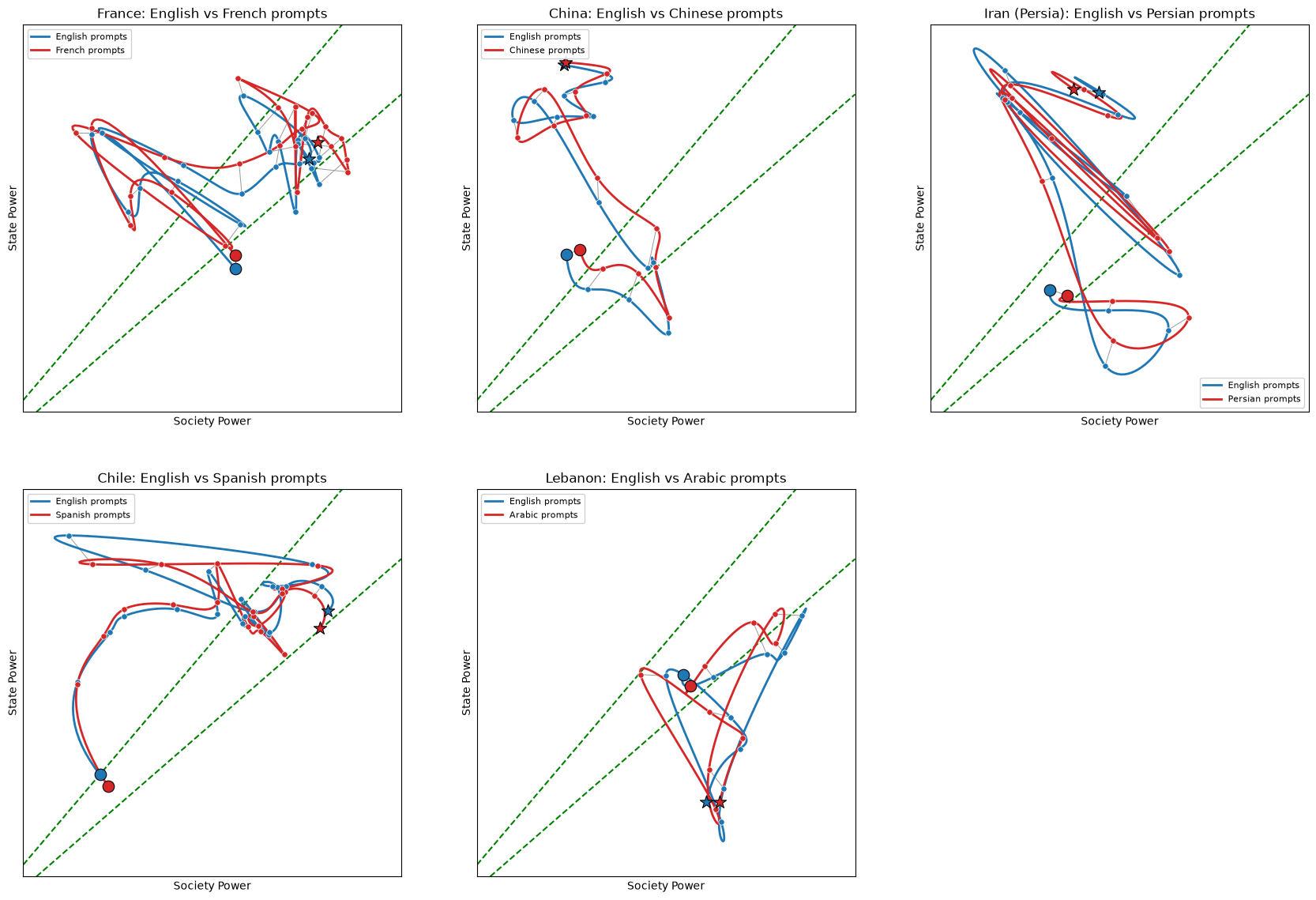}}%
    {\fbox{\parbox[c][3cm][c]{\textwidth}{\centering\footnotesize
       \textcolor{red}{\texttt{experiments/results\_lang/lang-ablation.png}}\\[2pt]
       (run \texttt{lang\_ablation.py})}}}
  \caption{Prompt-language ablation. Each panel overlays a country's ensemble-mean
  trajectory scored with English prompts (blue) and with prompts translated into
  the country's own language (red): French/France, Chinese/China, Persian/Iran,
  Spanish/Chile, Arabic/Lebanon.
  Circles mark the start period, stars the last; thin gray connectors show each
  period's displacement when the language changes. Only the prompt prose is
  translated; the rubric and the (English) output fields are held fixed, so the
  two paths are directly comparable.}
  \label{fig:language}
\end{figure*}

Because the pipeline prompts in English throughout, a natural worry is that the
scores partly reflect the language of the prompt rather than the country's
history, given that the models' English and non-English knowledge is drawn from
different, unevenly sized corpora. We probe this directly: we re-ran the full
four-model ensemble with every prompt (including the country's name) translated
into the language most tied to the country in question, French for France,
Chinese for China, Persian for Iran, Spanish for Chile, and Arabic for Lebanon,
keeping the rubric, the two-call structure, and the requested (English) output
fields identical, so only the prose language changes.
Figure~\ref{fig:language} overlays each country's English- and native-language
ensemble trajectory. The paths track each other closely: the root-mean-square
per-period displacement between the two language versions is $0.3$--$0.5$ rubric
points (largest for Iran, $0.53$), and the end-of-trajectory ($2020$) positions
differ by $0.08$--$0.56$ points (mean $0.33$)---both small relative to the
cross-model spread of Section~\ref{sec:reliability}. First-difference rank
correlations between the two language versions are high on both axes
(society $\rho=0.86$--$0.99$, state $\rho=0.88$--$0.98$), so the \emph{timing and
direction} of the moves is largely language-invariant.

We read this as prompt language having little effect on the scores, and
therefore little effect on their accuracy: switching the prompt out of English
neither removes nor amplifies the trajectories, which suggests the effect of the
predominantly English, Western-leaning training corpus on \emph{these}
measurements is small. We caution, however, that this is not evidence the
underlying corpus bias is absent. The language a model is \emph{prompted} in and
the language it \emph{reasons} in need not coincide~\citep{shi2023multilingual,
wendler2024llamas}: multilingual models often reason best through English and
appear to use it as an internal pivot, and under our translated prompts the
models frequently produced their chain-of-thought and key-event summaries in
English regardless of the prompt language. Translating the prompt therefore
changes the interface, not necessarily the evidence base the scores rest on, and
does not by itself pull the model off its English- and Western-centric sources.
The residual shifts are thus a lower bound on, not a full audit of, language
sensitivity (translation quality and the five-country scope also bound the
claim), but they do suggest the trajectories are driven more by the historical
record the models encode than by the surface language of the prompt.

\subsection{Qualitative case studies}
\textbf{Iran, 1979.} All four models place a society peak in the 1970s that
collapses into state dominance in the 1980s (the ensemble society score falls
sharply while state climbs past 8). This matches the book's account of the
revolution: the mobilization that toppled the Shah briefly raised society's
power, but the new revolutionary state then consolidated and demobilized that
society rather than balancing it, ending in a Despotic Leviathan. \textbf{Chile,
1973/1990.} The 1973 coup is the largest single move (society crashing, state
spiking near the ceiling), followed by recovery after the 1990 plebiscite, a
country leaving and re-entering the corridor. \textbf{United States.} Like the UK, the US traces a long co-growth of state and
society, ending inside the corridor as a Shackled Leviathan, with the Civil War
its largest single move. It sits a little higher on the state axis than the UK,
reflecting the twentieth-century build-out of the federal and security state, but
it does not cross into state dominance. The models agree least here
(Section~\ref{sec:reliability}), differing on how strong that central state grew
relative to civil society, yet they concur on the corridor placement, so the US
reads as a second Shackled case alongside the UK rather than an exception.

\section{Discussion: LLM vs.\ Human Researcher or Expert Panel}
\label{sec:discussion}

Both an LLM and a human expert (or panel) performing this coding are biased; the
useful question is \emph{how the bias profiles differ}, and which task
properties favor which coder.

\paragraph{What a human panel brings.} Domain expertise, sensitivity to
contested historiography, the ability to flag ``this case is genuinely
disputed,'' and accountable, attributable judgment. Established indices such as
V-Dem manage individual coder bias with multiple experts per case and a latent
measurement model that estimates and corrects for coder
disagreement~\citep{pemstein2018vdem}; content analysis formalizes this as
inter-coder reliability~\citep{krippendorff2018content}. The costs are equally
real: expert panels are slow, expensive, hard to scale to arbitrary
country-periods, and carry their own systematic biases (national, ideological,
disciplinary) that are difficult to audit because they live in individual
judgment.

\paragraph{What an LLM changes.} An LLM annotator is fast, cheap at scale,
and \emph{transparent in its instructions}: the rubric and prompts are explicit
artifacts that can be inspected, criticized, versioned, and re-run identically.
To put a number on ``cheap,'' the entire twelve-country, four-model sweep in this
paper (on the order of $1{,}600$ scoring and event calls) cost roughly \$15 in
API charges and ran in a few hours on one laptop; an expert-panel coding of
the same country-periods would cost orders of magnitude more in time and money.
Its bias, however, is \emph{opaque in origin}: it reflects the distribution of
its training data~\citep{bender2021dangers,santurkar2023whose,feng2023pretraining},
which over-represents English-language, recent, and Western sources, and it tends
toward received consensus, likely flattening contested or under-documented
episodes rather than flagging them. It has no accountable author, and its scores
can shift with model version or provider (Section~\ref{sec:reliability}).

\paragraph{Reliability as the bridge.} The inter-coder logic that human panels
use for humans applies to models: agreement across independent models is a
measurable proxy for reliability. Where independent models agree on the moves, a
score is more likely to reflect a shared signal than one model's idiosyncrasy;
where they disagree, the disagreement usefully localizes the contested cases a
human panel would also flag. Reliability is necessary but not sufficient for
validity, though: models trained on overlapping corpora can agree on a shared
bias, which is why we treat both inter-model agreement and V-Dem consistency
as evidence about \emph{reliability and consensus}, not accuracy.

\paragraph{When to use which.} We argue the LLM pipeline is best positioned as a
\emph{scalable first pass and a reproducible baseline} (generating candidate
trajectories over many countries cheaply, surfacing where models disagree, and
freeing expert attention for the contested cases) rather than as a replacement
for expert coding where accountability and contested-case judgment are
paramount.

\section{Limitations}
\label{sec:limitations}
Several limitations bound what this proof of concept establishes.
\emph{Circularity of the consistency check.} V-Dem and similar indices are
almost certainly in the models' training data, so agreement with them cannot
distinguish measuring the construct from recalling memorized values; we therefore
frame the check as concurrent consistency, not accuracy, and disentangling the
two (e.g., counterfactual-history probes, or ablating the events step) is future
work. \emph{Construct compression.} The two axes reduce vast, contested
constructs to single numbers, and our single state-capacity indicator only
partially aligns with the theory's notion of state power. \emph{Coverage and
language bias.} Training corpora over-represent English-language, recent, and
Western sources, weakest for exactly the cases we most need to read carefully
(Iran, China, Chile, Colombia, and the Congo), and we prompt in English
throughout; our prompt-language probe (Section~\ref{sec:language}) finds the
trajectories largely stable when the prompts are translated, but it covers only
five countries and cannot rule out shared cross-lingual bias.
\emph{Path dependence.} In-context anchoring
can propagate an early mis-score through the whole trajectory; we do not ablate
it here. \emph{Model idiosyncrasy.} We report the mean-across-models atlas
(Fig.~\ref{fig:atlas-mean}) as the headline reading and quantify per-axis spread
(Section~\ref{sec:reliability}); still, all four are English-centric LLMs and may
share bias. \emph{Scale and statistics.} Twelve countries and short per-country
series preclude cross-national generalization or strong significance claims. We
deliberately keep the study small; the natural next steps, a purpose-built human
coding of this exact task (beyond the V-Dem comparison we already report),
prompt and period-granularity ablations, a broader non-English prompting probe
(beyond the five languages in Section~\ref{sec:language}), and more
expert indices, are left to future work.

\section{Conclusion}
We presented, as a proof of concept, a method and a provider-agnostic tool that
operationalize the Narrow Corridor's (state, society) space into reproducible
trajectories, together with a protocol for checking their consistency with
expert consensus and their reliability across models, and released it as a
one-command exercise. The resulting twelve-country atlas is a first
quantified rendering of a framework that has been purely qualitative, and the
accompanying consistency check and bias discussion frame honestly what such
LLM-generated measurements are, and are not, good for.

\section*{Reproducibility}
All code, prompts, the on-disk response cache, and every run (JSON, static
plots, animations) are released at
\url{https://github.com/esonghori/narrow-corridor-llm}. The exact
experiments in this paper are reproduced by
\texttt{paper/experiments/run\_experiments.py} and \texttt{analysis.py}; see
\texttt{paper/README.md}. The interactive gallery of animated trajectories is
hosted at \url{https://esonghori.com/narrow-corridor-llm/gallery/}.

\bibliographystyle{plainnat}
\bibliography{references}

\appendix
\onecolumn
\section{Example: one period's full prompts and responses}
\label{app:example}
To make the two-call-per-period pipeline concrete, below is the \emph{verbatim}
exchange for a single period, the United Kingdom in 1829--1838, under
\texttt{claude-opus-4-8}: the chain-of-thought events prompt and the model's
reply, then the scoring prompt (which embeds the shared $0$--$10$ rubric of
Section~\ref{sec:method} and the trajectory so far) and the schema-validated JSON
score it returns.

\begin{lstlisting}
======================================================================
(1) EVENTS-AND-TRENDS PROMPT  [chain-of-thought, step 1]
======================================================================
Considering Daron Acemoglu and James A. Robinson's framework in *The Narrow
Corridor: States, Societies, and the Fate of Liberty*, identify the major
historical events and trends in United Kingdom during 1829-1838 that materially
affected EITHER the power of the state OR the power of civil society.

Be concrete and specific to 1829-1838. Distinguish slow-moving trends
(institutional, economic, demographic) from discrete events. For each item, note
briefly whether it strengthened or weakened the state, society, or both.

Respond with two short bulleted lists:

Trends:
- ...

Events:
- ...

======================================================================
(2) MODEL RESPONSE  [events and trends]
======================================================================
# UK, 1829-1838: State vs. Society in the Acemoglu-Robinson Framework

**Trends:**

- **Industrialization and urbanization**: The continued growth of manufacturing towns (Manchester, Birmingham, Leeds) concentrated new populations and created an assertive industrial middle class and urban working class--**strengthened society** by generating new mobilizing groups and demands for representation.

- **Growth of the popular press and associational life**: Expanding newspapers, political unions, and pamphleteering deepened the capacity for collective action and public opinion--**strengthened society**.

- **Fiscal-administrative modernization**: The gradual professionalization of governance (culminating in reforms like centralized poor-law administration) expanded the state's administrative reach and bureaucratic capacity--**strengthened the state**.

- **Rising economic distress among agricultural and industrial laborers**: Falling wages and mechanization fueled discontent, feeding movements like the emerging Chartist and trade-union agitation--**strengthened society's mobilization**, while pressuring the state.

- **Erosion of the old confessional/aristocratic settlement**: The long decline of exclusive Anglican-aristocratic monopoly on politics set the stage for enfranchisement of new groups--**shifted power toward society**.

**Events:**

- **Catholic Emancipation Act (1829)**: Removed most civil disabilities on Catholics, following mass mobilization (O'Connell's Catholic Association)--**strengthened society** (demonstrated concessions could be extracted) while showing the state adapting.

- **Metropolitan Police Act (1829)**: Peel's creation of the London police established a permanent, centralized coercive apparatus--**strengthened the state** (enhanced monopoly on order).

- **Swing Riots (1830-1831)**: Agrarian uprisings against mechanization and low wages; suppressed by state force but signaled rural society's capacity for disruption--**mobilized society**, prompted **state repression**.

- **Great Reform Act (1832)**: Redistributed seats to industrial towns and expanded the franchise to more middle-class men after intense extra-parliamentary pressure--**strengthened society** by broadening representation, though it excluded the working class.

- **Abolition of Slavery Act (1833)**: Product of sustained civil-society campaigning--**strengthened society's** demonstrated influence over policy.

- **Factory Act (1833)**: Regulated child labor and introduced inspectors--**strengthened both**: the state's regulatory reach and society's protective claims.

- **Poor Law Amendment Act (1834)**: Centralized and disciplined poor relief via workhouses and a national commission--**strengthened the state's** administrative power, provoking societal backlash.

- **Municipal Corporations Act (1835)**: Reformed and standardized local government, opening town councils to elected ratepayers--**strengthened both** local state capacity and civic participation.

- **Publication of the People's Charter (1838)**: Launched Chartism, articulating working-class demands (universal male suffrage, secret ballot)--**strengthened society's** organized mass mobilization.

*Caveat: Acemoglu and Robinson use this era as a case of Britain moving within "the corridor"--balanced growth of state and society through contested reform rather than either overwhelming the other.*

======================================================================
(3) SCORING PROMPT  [step 2; embeds the shared rubric + trajectory so far]
======================================================================
You are placing the country of United Kingdom in the 2D space of Acemoglu and
Robinson's *The Narrow Corridor*: the x-axis is the power of civil society and
the y-axis is the power of the state. We are tracing the path from 1789
to 2020.

Use this fixed 0-10 rubric so values are comparable across periods and countries.

STATE POWER -- the capacity of the state to enforce its will, administer
territory, collect taxes, and project authority:
  0-2  collapsed / nonexistent: no effective central authority.
  3-4  weak: limited reach, contested control, depends on local powerbrokers.
  5-6  moderate: functioning bureaucracy and military with real but bounded reach.
  7-8  strong: pervasive administration, effective coercion and revenue capacity.
  9-10 overwhelming: near-total penetration of society; little it cannot enforce.

SOCIETY POWER -- the capacity of civil society to organize, mobilize, hold the
state accountable, and resist domination (norms, associations, press, protest,
independent institutions):
  0-2  atomized: no autonomous organization; collective action near-impossible.
  3-4  weak: sporadic, easily suppressed mobilization.
  5-6  moderate: real associational life and periodic effective mobilization.
  7-8  strong: dense civic institutions routinely constraining the state.
  9-10 dominant: society pervasively shapes and checks state action.

Guidance for objectivity:
- Judge each period on its own terms; do not assume the country trends toward
  liberty, autocracy, or any particular corner of the space.
- Weigh evidence both for and against a shift before settling on a number.
- Avoid presentism: score by the standards and information of the period, not
  by how things turned out later.
- These two axes are independent: a strong state can coexist with a strong or a
  weak society. Score them separately.

Score the period 1829-1838, building on the established trajectory. Start
from where United Kingdom stood at the end of 1819-1828, then apply the
changes implied by the events and trends below. Report both the per-axis change
and the resulting absolute values (anchored to the rubric, not just to the prior
point). A period with little real change should show changes near 0.0.

Historical events and trends in 1829-1838:
# UK, 1829-1838: State vs. Society in the Acemoglu-Robinson Framework

**Trends:**

- **Industrialization and urbanization**: The continued growth of manufacturing towns (Manchester, Birmingham, Leeds) concentrated new populations and created an assertive industrial middle class and urban working class--**strengthened society** by generating new mobilizing groups and demands for representation.

- **Growth of the popular press and associational life**: Expanding newspapers, political unions, and pamphleteering deepened the capacity for collective action and public opinion--**strengthened society**.

- **Fiscal-administrative modernization**: The gradual professionalization of governance (culminating in reforms like centralized poor-law administration) expanded the state's administrative reach and bureaucratic capacity--**strengthened the state**.

- **Rising economic distress among agricultural and industrial laborers**: Falling wages and mechanization fueled discontent, feeding movements like the emerging Chartist and trade-union agitation--**strengthened society's mobilization**, while pressuring the state.

- **Erosion of the old confessional/aristocratic settlement**: The long decline of exclusive Anglican-aristocratic monopoly on politics set the stage for enfranchisement of new groups--**shifted power toward society**.

**Events:**

- **Catholic Emancipation Act (1829)**: Removed most civil disabilities on Catholics, following mass mobilization (O'Connell's Catholic Association)--**strengthened society** (demonstrated concessions could be extracted) while showing the state adapting.

- **Metropolitan Police Act (1829)**: Peel's creation of the London police established a permanent, centralized coercive apparatus--**strengthened the state** (enhanced monopoly on order).

- **Swing Riots (1830-1831)**: Agrarian uprisings against mechanization and low wages; suppressed by state force but signaled rural society's capacity for disruption--**mobilized society**, prompted **state repression**.

- **Great Reform Act (1832)**: Redistributed seats to industrial towns and expanded the franchise to more middle-class men after intense extra-parliamentary pressure--**strengthened society** by broadening representation, though it excluded the working class.

- **Abolition of Slavery Act (1833)**: Product of sustained civil-society campaigning--**strengthened society's** demonstrated influence over policy.

- **Factory Act (1833)**: Regulated child labor and introduced inspectors--**strengthened both**: the state's regulatory reach and society's protective claims.

- **Poor Law Amendment Act (1834)**: Centralized and disciplined poor relief via workhouses and a national commission--**strengthened the state's** administrative power, provoking societal backlash.

- **Municipal Corporations Act (1835)**: Reformed and standardized local government, opening town councils to elected ratepayers--**strengthened both** local state capacity and civic participation.

- **Publication of the People's Charter (1838)**: Launched Chartism, articulating working-class demands (universal male suffrage, secret ballot)--**strengthened society's** organized mass mobilization.

*Caveat: Acemoglu and Robinson use this era as a case of Britain moving within "the corridor"--balanced growth of state and society through contested reform rather than either overwhelming the other.*

Trajectory so far (period: <society_power, state_power>):
1789-1798: <6.0, 6.5>
1799-1808: <5.5, 7.5>
1809-1818: <5.5, 7.5>
1819-1828: <6.5, 7.5>

Return a JSON object with exactly these fields:
- society_power (float, 0-10): absolute society power at the end of 1829-1838
- state_power (float, 0-10): absolute state power at the end of 1829-1838
- society_change (float): society_power change vs. 1819-1828
- state_change (float): state_power change vs. 1819-1828
- key_event (string, <= ~10 words: the defining event/trend driving the change)
- reasoning (string, 1-2 sentences justifying the values)

======================================================================
(4) MODEL RESPONSE  [validated against the PeriodScore JSON schema]
======================================================================
{"society_power": 7.5, "state_power": 8.0, "society_change": 1.0, "state_change": 0.5, "key_event": "Great Reform Act (1832) and rise of mass mobilization", "reasoning": "Society surged as Catholic Emancipation, the Reform Act, abolition, and Chartism demonstrated dense associational life extracting concessions, while the state modernized via the Metropolitan Police, centralized Poor Law, and factory inspection\u2014balanced growth within the corridor."}
\end{lstlisting}

\end{document}